\documentclass[showpacs,amsmath,amssymb,pra,superscriptaddress]{revtex4}

\usepackage{graphicx}
\usepackage{dcolumn}
\usepackage{bm}

\begin{document}


\title{Electronic Structure of Atoms in Magnetic Quadrupole Traps}
\date{\today}
\pacs{32.80.Pj,31.15.Ar,32.10.Dk,32.60.+i}

\author{Igor Lesanovsky}
\email[]{ilesanov@physi.uni-heidelberg.de}
\affiliation{%
Physikalisches Institut, Universit\"at Heidelberg, Philosophenweg 12, 69120 Heidelberg, Germany}%
\author{J\"org Schmiedmayer}
\email[]{joerg.schmiedmayer@physi.uni-heidelberg.de}
\affiliation{%
Physikalisches Institut, Universit\"at Heidelberg, Philosophenweg 12, 69120 Heidelberg, Germany}%
\author{Peter Schmelcher$^{~\S,}$}
\email[]{Peter.Schmelcher@pci.uni-heidelberg.de}
\affiliation{%
Physikalisches Institut, Universit\"at Heidelberg, Philosophenweg 12, 69120 Heidelberg, Germany}%
\affiliation{%
Theoretische Chemie, Institut f\"ur Physikalische Chemie,
Universit\"at Heidelberg,
INF 229, 69120 Heidelberg, Germany}%
\thanks{Corresponding author}

\date{\today}

\begin{abstract}\label{txt:abstract}
We investigate the electronic structure and properties of atoms
exposed to a magnetic quadrupole field. The spin-spatial as
well as generalized time reversal symmetries are established and
shown to lead to a two-fold degeneracy of the electronic states
in the presence of the field. Low-lying as well as highly excited
Rydberg states are computed and analyzed for a broad regime of field gradients.
The delicate interplay between the Coulomb and various magnetic
interactions leads to complex patterns of the spatial spin
polarization of individual excited states. Electromagnetic
transitions in the quadrupole field are studied in detail thereby
providing the selection rules and in particular the transition wavelengths
and corresponding dipole strengths.  The peculiar property that the quadrupole
magnetic field induces permanent electric dipole moments of the atoms
is derived and discussed.
\end{abstract}

\maketitle
%
\section{Introduction}\label{sec:introduction}
%

External fields represent an excellent tool to influence and
control the motion of quantum systems and in particular to
generate new, sometimes very surprising, properties of e.g. atoms
and molecules. In the case of static homogeneous magnetic fields
the interest in the electronic structure of the hydrogen atom
reached its maximum in the late 1980's \cite{Friedrich89}. This
was motivated on the one hand by the experimental accessibility of
highly excited Rydberg states via laser spectroscopy as well as
the astronomical observation of spectra from magnetic white dwarfs
possessing a hydrogen-rich atmosphere \cite{Ruder94}. On the other
hand the development of new numerical techniques allowed to
address the regime where the competition of the Coulomb and
diamagnetic interaction leads to unusual and complex properties
and phenomena. Besides the vivid interest in magnetized structures
the hydrogen atom served as a paradigm of a nonseparable and
nonintegrable system possessing major impact with respect to the
development of several fields such as quantum chaos, semiclassics
of nonintegrable systems and nonlinear dynamics in general (see
\cite{Friedrich89,Friedrich97,Schmelcher98,Schmelcher97} and
references therein).

In contrast to the case of a homogeneous magnetic field there
exist no investigations on the electronic structure and properties
of atoms in inhomogeneous or trapping magnetic field
configurations. Apart from the fact that this problem is of
fundamental interest, experimental techniques allow nowadays to
manufacture magnetic microtraps on atom chips \cite{Folman02}. The
latter consists of surface-mounted structures carrying microscopic
charges and/or currents which create a miniature landscape of
inhomogeneous electric and/or magnetic fields. In particular it is
possible to build tight magnetic traps which involve magnetic
field configurations with large field gradients up to
${\mathcal{B}} \approx 10^8 \frac{G}{cm}$. Since these traps
strongly confine the atoms it is to be expected that the
electronic structure of excited or highly excited Rydberg states
is influenced significantly by the spatially varying field.
Quadrupole magnetic fields represent a key configuration which is
used in many setups of traps. We therefore investigate here the
electronic structure of atoms exposed to a quadrupole magnetic
field (see also ref.\cite{Lesanovsky04}).

In detail we proceed as follows. In section \ref{sec:hamiltonian}
we present the electronic Hamiltonian of the atom in the
quadrupole field including the interaction of the spatial as well
as spin degrees of freedom with the external field. Section
\ref{sec:symmetries} contains a discussion of the remarkable
spin-spatial symmetries of the Hamiltonian including unitary as
well as antiunitary symmetries and the related constants of
motion. As a result we encounter a two-fold degeneracy of each
eigenstate. Section \ref{sec:j_z_eigenstates} discusses the
eigenstates obeying different symmetries and the resulting
consequences for expectation values of observables. Section
\ref{sec:computational_approach} briefly outlines our
computational approach. In section \ref{sec:results} we present
the results of our numerical investigations establishing new
spectral and other properties for low-lying and highly excited
states for weak and strong gradients. The spin expectation values
and in particular the spatial distribution of the spin polarization of the excited
states are studied in detail. Electromagnetic transitions in the
quadrupole field including dipole strengths and wavelengths are
given too. Whereever appropriate a comparison with the case of a
homogeneous magnetic field is performed. Finally the peculiar
property of magnetic field-induced permanent electric dipole
moments is derived and investigated in depth. Section
\ref{sec:conclusion_outlook} contains the conclusions and outlook
as well as some discussion of relevant aspects going beyond the
present investigation.

%
\section{The Hamiltonian}\label{sec:hamiltonian}
%
In ultracold atomic physics moving atoms in inhomogeneous magnetic
fields are considered to be neutral point-like particles which
couple only through their total angular momentum to the external
field \cite{Folman02,Bergeman89}. In slowly varying magnetic fields it is
appropriate and convenient to describe the dynamics of the atoms by
means of an adiabatic approximation, where one assumes that the
magnetic moment is oriented either parallel or antiparallel to the
magnetic field. In this way the interaction potential becomes
proportional to the modulus of the magnetic field. As long as the
spatial extent of the atom is much less than the typical
variations of the field and if one is only interested in the
center of mass motion of the atom the adiabatic approximation can
be applied reliably.

In the present investigation we are interested in the electronic
structure of excited atoms exposed to a quadrupole magnetic trap
or field. Therefore we have to go beyond the above approximation
in the sense that the detailed coupling of both the electronic
spatial as well as the spin degrees of freedom to the external
field is taken into account. The majority of todays experiments on
ultracold atoms deal with alkali-atoms \cite{Folman02}.
Alkali-atoms become rapidly hydrogen-like if their single valence
electron is excited. The detailed structure of the core becomes
increasingly irrelevant with increasing degree of excitation. We
therefore assume that the outer electron is exposed to the field
of a single positive point charge. Furthermore we neglect the
spin-orbit ($LS$) coupling and the coupling of the nuclear and
electronic spin, i.e. the hyperfine interaction. These
interactions show an $r^{-3}$-dependence, i.e. a rapid decay with
increasing $r$. Thus for sufficient highly excited states their
contributions are negligible compared to those of the
Coulomb-potential and the interaction with the external field.
Nevertheless one can discuss how the inclusion of spin-orbit
coupling might modify the properties of the system. For hydrogen
in a homogeneous field this relativistic effect changes the
symmetries of the system crucially: $S_z$ and $L_z$ are no longer
conserved separately and only $J_z$ remains a conserved quantity.
The result is a restructuring of the energy spectrum since
previously separated symmetry sub-spaces become coupled. In
particular this manifests itself in the emergence of avoided
crossing \cite{Ruder94}. For the atom in the quadrupole field such
dramatic effects are not expected since the additional
$LS$-coupling term does not reduce the symmetry of the system.
Additionally, considering its $r^{-3}$-dependence, the spin-orbit
interaction can certainly be treated by means of perturbation
theory in the regime of spatially extensive and energetically
highly excited Rydberg states.

In the presence of a magnetic field the motion of the center of
mass (CM) and internal (electronic) degrees of freedom of an atom
do not decouple. This holds in particular for the case of a
homogeneous magnetic field
\cite{Lamb59,Avron78,Johnson83,Schmelcher94} and therefore has
also to be expected for an inhomogeneous field. However,
CM-motional effects on the electronic structure become only
significant in certain parameter and/or energetic regimes
\cite{Ruder94,Schmelcher92,Dippel94}. In order to approximately
decouple the CM and electronic dynamics we take advantage of the
heavy atomic mass compared to the electron mass ($m_A\gg m_e$).
Additionally we assume an ultracold atom whose CM motion takes
place on much larger time scales than the electron dynamics, even
for highly excited electronic states. The Hamiltonian describing
the electronic motion then reads
\begin{eqnarray}
H_C=\frac{1}{2m_e}\left(\vec{p}+e\vec{A}(\vec{r})\right)^2-\frac{e^2}{4\pi\epsilon_0|\vec{r}|}+\frac{g_s
\mu_B}{\hbar}\vec{S}\vec{B}(\vec{r}).
\label{eq:spinor_hamiltonian}
\end{eqnarray}
Where we have assumed that the atomic core (nucleus) is fixed at
the trap center which is the origin of the coordinate system. The
magnetic quadrupole field is given as
\begin{eqnarray}
\vec{B}(\vec{r})=b\left(%
\begin{array}{c}
  x \\
  y \\
  -2z \\
\end{array}%
\right)\label{eq:quadrupole_field}.
\end{eqnarray}
The gradient $b$ is the only parameter characterizing the
steepness of the quadrupole trap. The vector field is rotationally
symmetric around the $z$-axis and invariant under the $z$-parity
operation.
\begin{figure}[htb]\center
\includegraphics[angle=0,width=6.7cm]{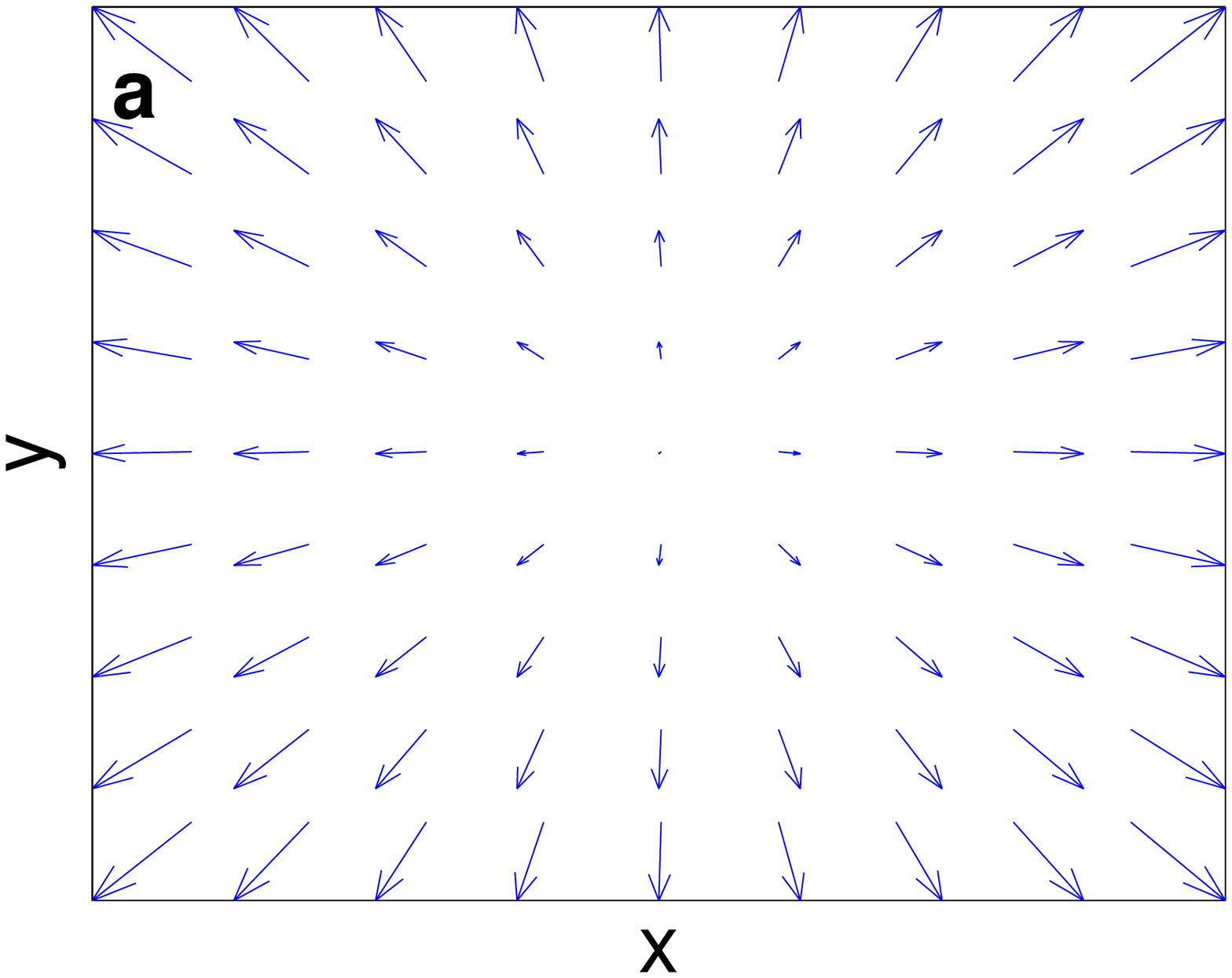}
\includegraphics[angle=0,width=6.7cm]{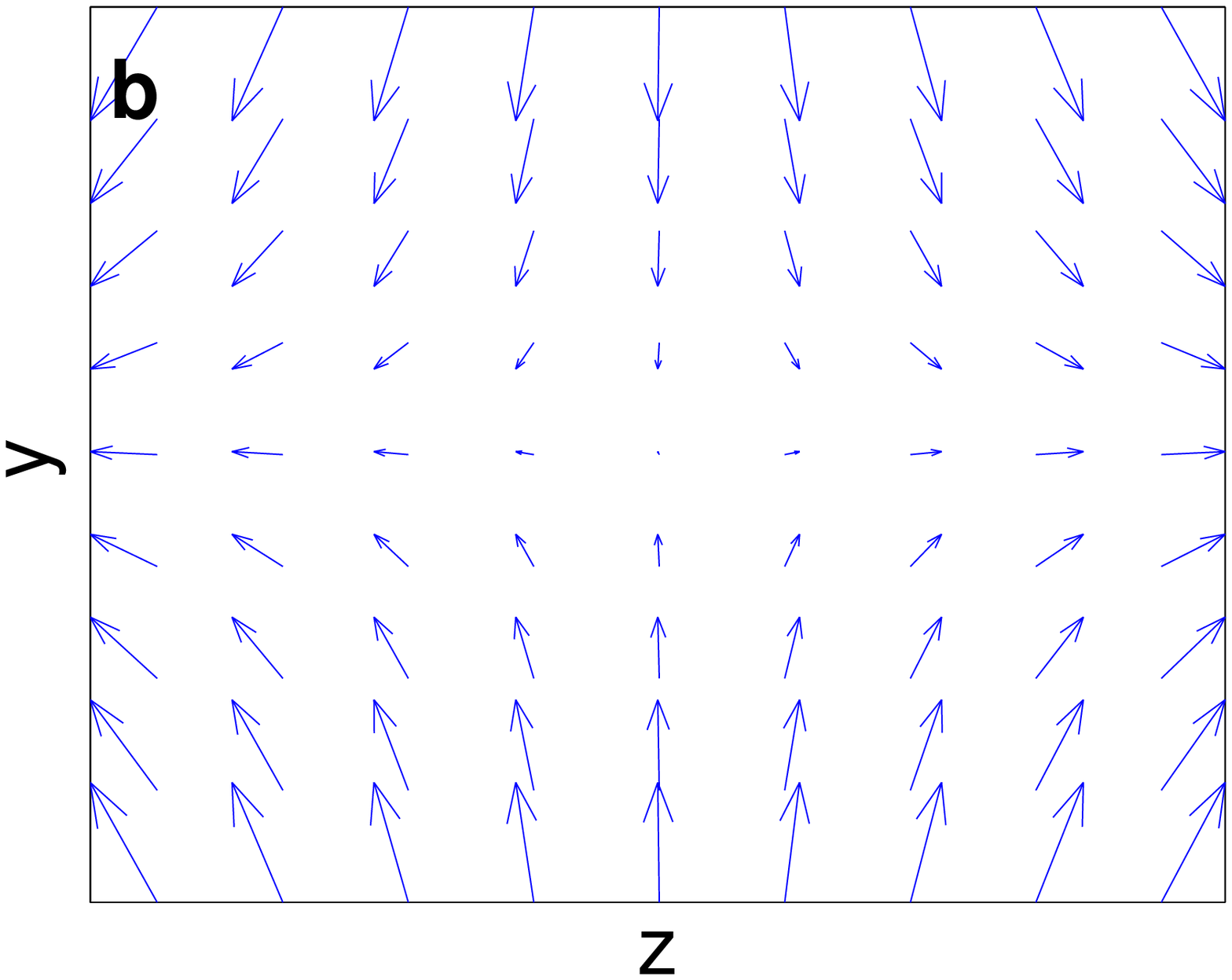}
\caption{Vectorial plots of the magnetic quadrupole field
(\ref{eq:quadrupole_field}). The intersections are the
$(x,y)$-plane for $z=0$ (\textbf{a}) and the $(y,z)$-plane for
$x=0$ (\textbf{b}).}\label{fig:quad_field}
\end{figure}
Figure \ref{fig:quad_field} shows intersections of this vector
field for $z=0$ and $x=0$, correspondingly. A vector potential
belonging to the quadrupole field reads
\begin{eqnarray}
\vec{A}(\vec{r})=\frac{1}{3}\left[\vec{B}(\vec{r})\times\vec{r}\right]=b\left(%
\begin{array}{c}
  yz \\
  -xz \\
  0 \\
\end{array}
\right),\label{eq:vector_potential}
\end{eqnarray}
which is a Coulomb gauge ($\nabla\vec{A}(\vec{r})=0$). In
Cartesian coordinates and adopting atomic units
\footnote{$\hbar=1$, $m_e=1$, $a_0=1$: The magnetic gradient unit
then becomes $b=1a.u.=4.44181\cdot 10^{15} \frac{T}{m}$. The magnetic
field strength unit is $B=1a.u.=2.35051\cdot 10^{5} T$} ($g_s=2$) the
spinor Hamiltonian (\ref{eq:spinor_hamiltonian}) becomes
\begin{eqnarray}
{H_C}=-\frac{1}{2}\triangle-\frac{1}{\sqrt{x^2+y^2+z^2}}
-b\,z{L}_z+\frac{b^2}{2}z^2\left(x^2+y^2\right)
+\frac{b}{2}\left(\sigma_xx+\sigma_yy-2\sigma_zz\right)\label{eq:hamiltonian_rc=0_cart}
\end{eqnarray}
with $L_z=-i\left(y\partial_x-x\partial_y\right)$ and $\sigma_x$,
$\sigma_y$, $\sigma_z$ being the Pauli spin matrices
($\vec{S}=\frac{1}{2}\vec{\sigma}$). The first two terms of $H_C$ in
equation (\ref{eq:hamiltonian_rc=0_cart}) represent the
Hamiltonian of the hydrogen atom without external fields. The
third term is the paramagnetic ($\propto b$) or Zeeman-term. In
contrast to the situation of the atom in a homogeneous field this
term now depends not only on $L_z$ but also linearly on the
$z$-coordinate. The fourth (diamagnetic) term ($\propto b^2$)
represents a quartic oscillator coupling term between the
cylindrical coordinates $\rho=\sqrt{x^2+y^2}$ and $z$. In a
homogeneous field the diamagnetic interaction is a pure harmonic
oscillator term proportional to $\rho^2$ and yields a confinement
perpendicular to the magnetic field. Finally the reader should
note that the coupling of the electronic spin to the quadrupole
field depends linearly on the Cartesian coordinates. The latter
prevents the factorization of the motions in the spin and spatial
degrees of freedom and renders the corresponding Schr{\"o}dinger
equation into a spinor equation. This is again in contrast to the
situation of a homogeneous magnetic field where the spin component
along the field is a conserved quantity.
%
%
\section{Symmetries}\label{sec:symmetries}
%
%
In this section we investigate the unitary as well as the
anti-unitary symmetries and conserved quantities of the
Hamiltonian (\ref{eq:hamiltonian_rc=0_cart}). We will then
construct the corresponding eigenfunctions respecting these
symmetries. This will help us (in Section
\ref{eq:jz_eigenfunction}) to derive properties of expectation
values. Some basic definitions and features of probability
densities are also introduced being utilized later to analyze
excited atomic states in the quadrupole field.

For the following discussion it is useful to transform the
Hamiltonian (\ref{eq:hamiltonian_rc=0_cart}) into spherical
coordinates. It reads
\begin{eqnarray}
{H_S}&=&-\frac{1}{2}\triangle_{r,\theta,\phi}-\frac{1}{r}+\frac{b^2}{2}r^4\cos^2\theta\sin^2\theta+\frac{b}{2}r\sin\theta\,K-b\,r\cos\theta\left({L}_z+\sigma_z\right)\label{eq:hamiltonian_rc=0_spher}
\end{eqnarray}
with $K$ being the matrix
\begin{eqnarray}
 K=\left(%
\begin{array}{cc}
  0 & e^{-i\phi} \\
  e^{i\phi} & 0 \\
\end{array}%
\right).
\end{eqnarray}
As a consequence of the rotational invariance of the quadrupole
field the $z$-component of the total angular momentum
$J_z=L_z+S_z$ is conserved, i.e. we have $\left[H_S,J_z\right]=0$.
Additionally we have the discrete symmetry represented by the
unitary operator $P_\phi OP_z$, i.e. $\left[H_s,P_\phi
OP_z\right]=0$. Here $O\equiv\sigma_x$ exchanges the components of
a $\frac{1}{2}$-spinor, $P_z$ is the $z$-parity operator
$P_z:z\rightarrow -z$ and $P_\phi$ represents the 'reflection'
$P_\phi:\phi\rightarrow 2\pi-\phi$. Apart from these symmetries
the Hamiltonian possesses two generalized anti-unitary time
reversal symmetries namely $TOP_z$ and $TP_\phi$. Both of them
involve the conventional time reversal operator $T$ ($T^2=1$),
which corresponds to complex conjugation. Applying the two
anti-unitary operators $TOP_z$ and $TP_\phi$ results in the
unitary operation $P_\phi O P_z$. The operators $TOP_z$, $TP_\phi$
and $P_\phi O P_z$ form an invariant Abelian sub-group. Together
with $J_z$ they obey the following (anti-)commutation rules:
\begin{eqnarray}
\left[J_z,TP_\phi\right]=\left\{J_z,TOP_z\right\}=\left\{J_z,P_\phi
OP_z\right\}=0\\
\left[TP_\phi,TOP_z\right]=\left[TP_\phi,P_\phi
OP_z\right]=\left[P_\phi
OP_z,TOP_z\right]=0\label{eq:top_pop_commutator}
\end{eqnarray}
Apparently the spin-spatial symmetry operations form a non-Abelian
symmetry group. This group generated by $P_\phi OP_z$ and $J_z$ is
isomorphic to $C_\infty \bigotimes C_s$.

In the following we will show, that the interplay of the above
symmetries leads to a degeneracy of the eigenvalues of the
Hamiltonian (\ref{eq:hamiltonian_rc=0_spher}). We consider a state
$\left|E,m\right>$ which is an eigenstate of the Hamiltonian
(\ref{eq:hamiltonian_rc=0_spher}) with the energy $E$ and to $J_z$
with the half-integer quantum number $m$ (see below). Since
$TOP_z$ commutes with $H_S$ the state $TOP_z\,\left|E,m\right>$ is
also an energy eigenstate with the energy $E$. By letting $J_z$
act on this state one obtains
\begin{eqnarray}
J_z\,TOP_z\,\left|E,m\right>=-TOP_z\,J_z\,\left|E,m\right>=-m\,TOP_z\,\left|E,m\right>.\nonumber
\end{eqnarray}
Thus the state $TOP_z\,\left|E,m\right>$ can be identified with
$\left|E,-m\right>$. Hence the states with the eigenvalues $m$ and
$-m$ are degenerate. This two-fold degeneracy of each energy level
in the presence of the inhomogeneous magnetic field is a
remarkable feature. It is reminiscent of the Kramers degeneracy of
spin $\frac{1}{2}$ systems in the absence of external fields
\cite{Haake01}.

In the case of an atom in a homogeneous magnetic field the
operators $L_z$, $P_z$ as well as parity $P$ and $T\sigma_z
P_\phi$ form the corresponding set of spatial and time reversal
symmetries if the field is oriented along the $z$-axis. Since
these operators commute they form an Abelian symmetry group
implying the well-known fact that there are no energy level
degeneracies in the homogeneous field.

In order to finalize this section we want to remark that most of
the above considerations hold not only for the case of a fixed
nucleus in the trap center but also for the full two-body problem.
If the dynamics of both the nucleus and the electron is considered
the conservation of $TP_\phi$, $TOP_z$ and $J_z$ of the two-body
system holds.
%
\section{$J_z$-, $TOP_z$- and $TP_\phi$-eigenstates}\label{sec:j_z_eigenstates}
%
Since the $J_z$ operator provides the eigenvalue equation
$J_z\left|m\right>=m\,\left|m\right>$ with half-integer $m$ the
spatial representation of the spinor eigenfunctions
$\left|m\right>$ are
\begin{eqnarray}
\left|m\right>=\left(%
\begin{array}{c}
  e^{i(m-\frac{1}{2})\phi} \\
  e^{i(m+\frac{1}{2})\phi} \\
\end{array}%
\right).\label{eq:jz_eigenfunction}
\end{eqnarray}
The eigenfunction $\left|m\right>$ can be used to reduce the
electron dynamics to a given $m$-subspace. The quantized motion of
the electron is then described by the Hamiltonian
$H_m=\left<m\right|H\left|m\right>$. We will exploit this fact
later when solving the stationary Schr\"odinger equation.

Eigenstates to the $TOP_z$-operator can be constructed by
superimposing two degenerate $J_z$-eigenstates. They read
\begin{eqnarray}
\left|E,\pm\right>^{TOP_z}=\frac{1}{\sqrt{2}}\left[\left|E,m\right>\pm
TOP_z\,\left|E,m\right>\right].\label{eq:TOP-state}
\end{eqnarray}
The corresponding eigenvalue relation is
\begin{eqnarray}
TOP_z\,\left|E,\pm\right>=\pm\left|E,\pm\right>.
\end{eqnarray}
Eigenfunctions of the anti-unitary operator $TP_\phi$ can be
constructed analogously:
\begin{eqnarray}
\left|E,\pm\right>^{TP_\phi}=\frac{1}{\sqrt{2}}\left[\left|E,m\right>\pm
TP_\phi\,\left|E,m\right>\right]\label{eq:TP-state}
\end{eqnarray}

The $J_z$- and $TOP_z$-eigenstates exhibit certain symmetry
properties. Exploiting such symmetries often turns out to be
extremely powerful and can save a lot of computational effort. It
is especially useful when one is interested in computing matrix
elements such as expectation values of certain observables.

The expectation value of an observable $M$ in an energy and $J_z$-
eigenstate $\left|E,m\right>$ is defined as
$\left<M\right>_{J_z}=\left<E,m\right|M\left|E,m\right>$ whereas
for a  $TOP_z$-eigenstate (\ref{eq:TOP-state}) it reads
\begin{eqnarray}
\left<M\right>^\pm_{TOP_z}&=&{^{TOP_z}}\left<E,\pm\right|M\left|E,\pm\right>^{TOP_z}\nonumber\\
&=&\frac{1}{2}\left[\left<M\right>_{J_z}+\left<TOP_z\,M\,TOP_z\right>_{J_z}\pm\left<TOP_z\,M\right>^*_{J_z}\pm\left<M\,TOP_z\right>_{J_z}\right].\label{eq:top_z_expectationvalue}
\end{eqnarray}\\
If $M$ satisfies the anti-commutation relation
$\left\{TOP_z,M\right\}=0$ the first two terms of
(\ref{eq:top_z_expectationvalue}) cancel each other and we obtain
\begin{eqnarray}
\left<M\right>^\pm_{TOP_z}=\pm\frac{1}{2}\left[\left<TOP_z\,M\right>^*_{J_z}-\left<TOP_z\,M\right>_{J_z}\right]
= \mp i\,\text{Im}\left(\left<TOP_z\,M\right>_{J_z}\right).
\end{eqnarray}
Since the expectation value of an observable is always a real
number it follows
\begin{eqnarray}
\left<M\right>^\pm_{TOP_z}&=&0,\label{eq:observable_property_1}
\end{eqnarray}
i.e. the expectation value of such an operator $M$ vanishes within
any $TOP_z$-eigenstate.

In a similar manner one can calculate the expectation value of an
observable $N$, which commutes with $J_z$ and $TOP_z$. Here
expression (\ref{eq:top_z_expectationvalue}) yields
\begin{eqnarray}
\left<N\right>^\pm_{TOP_z}=\left<N\right>_{J_z}\pm\left<N\,TOP_z\right>_{J_z}.
\end{eqnarray}
Exploiting  $\left[N,J_z\right]=0$ and using
$\left\{TOP_z,J_z\right\}=0$ one obtains
\begin{eqnarray}
\left<N\,TOP_z\right>_{J_z}&=&\frac{1}{m}\left<E,m\right|N\,TOP_z\,J_z\left|E,m\right>=-\frac{1}{m}\left<E,m\right|N\,J_z\,TOP_z\left|E,m\right>\nonumber\\
&=&-\frac{1}{m}\left<E,m\right|J_z\,N\,TOP_z\left|E,m\right>=-\left<N\,TOP_z\right>_{J_z}
\end{eqnarray}
Hence $\left<N\,TOP_z\right>_{J_z}=0$ leading to the final result
\begin{eqnarray}
\left<N\right>^\pm_{TOP_z}=\left<N\right>_{J_z},\label{eq:observable_property_2}
\end{eqnarray}
i.e. we obtain equal expectation values of $N$ for a $J_z$- or
$TOP_z$-eigenstate.

Finally we provide some relevant expressions for the spatial
probability density $W$ in the different representations. The
probability density belonging to the $J_z$-eigenstate
$\left|E,m\right>=\left|u\right>\left|\uparrow\right>+\left|d\right>\left|\downarrow\right>$
reads
\begin{eqnarray}
W_{J_z}\left(r,\theta,\phi\right)=r^2\sin\theta\left[\left|u\left(r,\theta\right)\right|^2+\left|d\left(r,\theta\right)\right|^2\right].
\end{eqnarray}
Since the whole $\phi$-dependence is contained in phase-factors
(see eq. (\ref{eq:jz_eigenfunction})) $W_{J_z}$ is invariant
under rotations around the $z$-axis. For the $TOP_z$-eigenstates
(\ref{eq:TOP-state}) one obtains
\begin{eqnarray}
W_{TOP_z}\left(r,\theta,\phi\right)&=&\frac{r^2\sin\theta}{2}\left[\left|u\left(r,\theta\right)\right|^2+\left|d\left(r,\theta\right)\right|^2+\left|u\left(r,\pi-\theta\right)\right|^2+\left|d\left(r,\pi-\theta\right)\right|^2\right.\\
&&\left.\pm
2\,\left\{u\left(r,\theta\right)d\left(r,\pi-\theta\right)+d\left(r,\theta\right)u\left(r,\pi-\theta\right)\right\}\cos\left(2m\,\phi\right)\right]\nonumber.
\end{eqnarray}
$W_{TOP_z}\left(r,\theta,\phi\right)$ is constructed from
$TOP_z$-eigenstates and obeys therefore the $TOP_z$-symmetry.
Since $W_{TOP_z}\left(r,\theta,\phi\right)$ is a real scalar
function the symmetries $T$ and $O$ are trivially satisfied
leading to
\begin{eqnarray}
W_{TOP_z}=TOP_z\,W_{TOP_z}\,(TOP_z)^{-1}=TP_z\,W_{TOP_z}\,(TP_z)^{-1}=P_z\,W_{TOP_z}\,(P_z)^{-1},
\end{eqnarray}
showing that $W_{TOP_z}\left(r,\theta,\phi\right)$ is invariant
under $z$-parity.

%
%
\section{Computational approach}\label{sec:computational_approach}
%
%
We employ the linear variational principle \cite{Szabo96} to find
the eigenvalues and eigenvectors of the Hamiltonian
(\ref{eq:hamiltonian_rc=0_spher}). The idea is to expand the exact
eigenfunctions of the Schr\"odinger equation in a complete basis
set of spinor orbitals that converge efficiently and accurately
towards the exact solution. Calculating the expansion coefficients
results in a large-scale algebraic eigenvalue
equation.\\
The basis set $\left\{B^\uparrow_{nl},
B^\downarrow_{\tilde{n}\tilde{l}}\right\}_m$ we apply for given
$m$ takes the following form for the upper and lower component, respectively:
\begin{eqnarray}
\left<r,\theta,\phi\mid B^\uparrow_{nl}\right>=
  R_{n}^{(\zeta,k)}(r)Y_l^{m-\frac{1}{2}}(\theta,\phi)\left|\uparrow\right>\quad,\quad
\left<r,\theta,\phi\mid B^\downarrow_{\tilde{n}\tilde{l}}\right>=
 R_{\tilde{n}}^{(\zeta,k)}(r)Y_{\tilde{l}}^{m+\frac{1}{2}}(\theta,\phi)\left|\downarrow\right>\label{eq:basis_set}
\end{eqnarray}
Here $Y_{l}^{m}(\theta,\phi)$ denote the spherical harmonics. Due
to the appearance of the upper and lower spinor orbitals for fixed
$m$ the energy eigenfunctions of the Schr\"odinger equation
computed with this basis set are a priori eigenfunctions of $J_z$.

The radial part of the orbitals used for the expansion reads
\begin{eqnarray}
R_{n}^{(\zeta,k)}(r)=\sqrt{\frac{n!}{(n+2k)!}}e^{-\frac{\zeta
r}{2}}(\zeta r)^{k}L_n^{2k}(\zeta r)\label{eq:radial_part}
\end{eqnarray}
with $L_n^{2k}(r)$ being the Laguerre polynomials. The parameters
$k$ and $\zeta$ can be tuned to optimize the convergence behavior
in different parts of the spectrum. $\zeta$ possesses the
dimension of an inverse length. In order to converge eigenstates
with the smallest possible number of basis functions $1/\zeta$ has
to be adapted for every energy regime such, that it corresponds to
the typical spatial extensions of the wavefunctions. For fixed $k$
and $\zeta$ the functions $R_{n}^{(\zeta,k)}(r)$ form a complete
functional set in $r$-space but are non-orthogonal leading to a
non-trivial overlap-matrix. The basis set (\ref{eq:basis_set}) is
complete in $r$-, $\theta$- and spin-space.

An eigenstate $\left|E,m\right>$ of the stationary Schr{\"o}dinger
equation can now be expanded in terms of the basis-functions
(\ref{eq:basis_set})
\begin{eqnarray}
\left|E,m\right>&=&\sum_{n=0,l=\left|m-\frac{1}{2}\right|}^{n<N,l<L}a_{n,l}\left|B^\uparrow_{nl}\right>+\sum_{\tilde{n}=0,\tilde{l}=
\left|m+\frac{1}{2}\right|}^{\tilde{n}<\tilde{N},\tilde{l}<\tilde{L}}b_{\tilde{n},\tilde{l}}\left|B^\downarrow_{\tilde{n}\tilde{l}}\right>\label{eq:expansion_of_state}
\end{eqnarray}
leading to the generalized spinor eigenvalue problem
$\mathbf{H}\vec{c}=E\mathbf{S}\vec{c}$, where $\mathbf{H}$ and
$\mathbf{S}$ are the corresponding matrix representation of the
Hamiltonian (\ref{eq:hamiltonian_rc=0_spher}) and the overlap
matrix, respectively:
\begin{eqnarray}
\mathbf{H}=\left(%
\begin{array}{cc}
  \left<B^\uparrow_{nl}\left|{H}\right|
B^\uparrow_{n^\prime l^\prime}\right> &
\left<B^\uparrow_{nl}\left|{H}\right|
B^\downarrow_{\tilde{n}^\prime \tilde{l}^\prime}\right> \\
  \left<B^\downarrow_{\tilde{n}\tilde{l}}\left|{H}\right|
B^\uparrow_{n^\prime l^\prime}\right> &
\left<B^\downarrow_{\tilde{n}\tilde{l}}\left|{H}\right|
B^\downarrow_{\tilde{n}^\prime \tilde{l}^\prime}\right> \\
\end{array}%
\right)\label{eq:hamiltonmatrix}\quad
\mathbf{S}=\left(%
\begin{array}{cc}
  \left<B^\uparrow_{nl}\mid B^\uparrow_{n^\prime l^\prime}\right> &
0 \\
  0 &
\left<B^\downarrow_{\tilde{n}\tilde{l}}\mid
B^\downarrow_{\tilde{n}^\prime \tilde{l}^\prime}\right> \\
\end{array}%
\right)\label{eq:overlapmatrix}
\end{eqnarray}
The vector $\vec{c}$ contains the expansion coefficients of
(\ref{eq:expansion_of_state}):
\begin{eqnarray}
\vec{c}=\left(%
\begin{array}{c}
  a_{n,l} \\
  b_{\tilde{n},\tilde{l}} \\
\end{array}%
\right)
\end{eqnarray}
The two off-diagonal blocks of $\mathbf{S}$ vanish due to the
orthogonality of the spin states $\left|\uparrow\right>$ and
$\left|\downarrow\right>$. This does not hold for $\mathbf{H}$
since the magnetic field leads to a coupling of the two spinor
components. With the basis set (\ref{eq:basis_set}) all entries of
the matrices (\ref{eq:hamiltonmatrix}) can be computed
analytically. To do this we have exploited recurrence identities
for the spherical harmonics and the Laguerre polynomials
\cite{Abramowitz68}. The matrices $\mathbf{H}$ and $\mathbf{S}$
possess a particular sparse appearance, e.g. $\mathbf{S}$ is
penta-diagonal, enabling us to go to large basis set dimensions.

In order to solve the above generalized eigenvalue problem we
employ the Arnoldi-method \cite{Sorensen} which is a so-called
Krylov-space method. This approach is especially suited for
eigenvalue equations involving large sparse matrices. The
Arnoldi-Method is perfectly suited to provide low-lying excited
states with fairly small basis sets. To calculate highly excited
states without computing the lower excited states we use the so
called shift-and-invert method. Here the generalized eigenvalue
problem is transformed to $
\left(\mathbf{H}-\sigma\,\mathbf{S}\right)^{-1}\mathbf{S}\vec{c^\prime}=\lambda\vec{c^\prime}$.
The eigenvalues $E$ of the original eigenvalue equation are
connected to $\lambda$ according to
$E_i=\sigma+\frac{1}{\lambda}$. Thus eigenvalues lying close to
the shift $\sigma$ become the largest eigenvalues in magnitude of
the operator
$\left(\mathbf{H}-\sigma\,\mathbf{S}\right)^{-1}\mathbf{S}$. Using
the Arnoldi-method these eigenvalues lying in a predefined energy
range of the spectrum will converge first. The shift-and-invert
method allows in principle to compute arbitrary parts of the
spectrum of the Hamiltonian ($\ref{eq:hamiltonian_rc=0_spher}$) by
resetting the shift $\sigma$ in successive calculations. Thereby
the basis parameter $\zeta$ has to be adapted. A good choice is
$\zeta=\frac{2\sqrt{\left|E\right|}}{0.775}$ where $E$ is the
lower boundary of the corresponding energy range. We remark that
there is no need to change the parameter $k$, which we always put
to zero.

Due to the Hylleraas-Undheim theorem \cite{Newton82} the numerical
approximate eigenvalues provide an upper bound for the exact
eigenvalues. In order to check the convergence behavior we
calculate a number of eigenvalues $E_i^{G_1}$ for a given basis
set size $G_1$. In a second step the size of the basis set is
increased significantly to $G_2$ and the same eigenvalues are
calculated again being now denoted by $E_i^{G_2}$. As a measure of
convergence we define the quantity
\begin{eqnarray}
K_i=\left|\frac{E_i^{G_1}-E_i^{G_2}}{E_i^{G_1}-E_{i-1}^{G_1}}\right|
\end{eqnarray}
where the difference of the same eigenvalue for the two basis
sizes $G_1$ and $G_2$ is divided by the distance to the lower
neighboring eigenvalue. For $K_i\le 0.01$ we assume the eigenvalue
$E_i^{G_2}$ to be well converged.

The maximum dimension of the basis sets we employed was
approximately 17000.  We thereby were able to converge several
thousand eigenstates and eigenvalues up to energies corresponding
to a hydrogen principle quantum number of $n\approx 60$ for weak
gradients and gradients $b$ ranging from $10^{-10}$ to $10$
covering the $J_z$-quantum numbers
$m\,\epsilon\,\left[-\frac{7}{2},\frac{7}{2}\right]$.
%
%
\section{Results}\label{sec:results}
%
%
In the following we discuss and analyze our results on the atom in
the quadrupole field. We address for different regimes of the
field gradient the effects of the quadrupole field on the spatial
appearance of the wavefunctions as well as their spin properties.
Furthermore we provide the selection rules for electric dipole
transitions. A comparison with properties of the atom in a
homogeneous magnetic field is performed to illuminate the specific
features in a quadrupole field.
%
%
\subsection{Spectral Properties}\label{subsec:spectral_properties}
%
%
Depending on the gradient $b$ of the quadrupole field one can
distinguish three regimes: the weak, the intermediate and the
strong gradient regime. In each of these regimes the atoms
exhibits unique features. The notion of a weak/strong gradient
does not refer to an absolute field gradient but depends on the
degree of excitation of the atom. It is defined to be the regime,
for which the magnetic compared to the Coulomb interaction is
weak/strong. For weak gradients the level structure is dominated
by the spin as well as the spatial Zeeman-term both of which
depend linearly on the field gradient. The level degeneracies at
$b=0$ are lifted and a linear splitting of the $n$-multiplets is
observed. There is no overlap between neighboring $n$-multiplets
which renders $n$ an approximately good quantum number.
\begin{figure}[htbp]\center
\includegraphics[angle=-90,width=7.1cm]{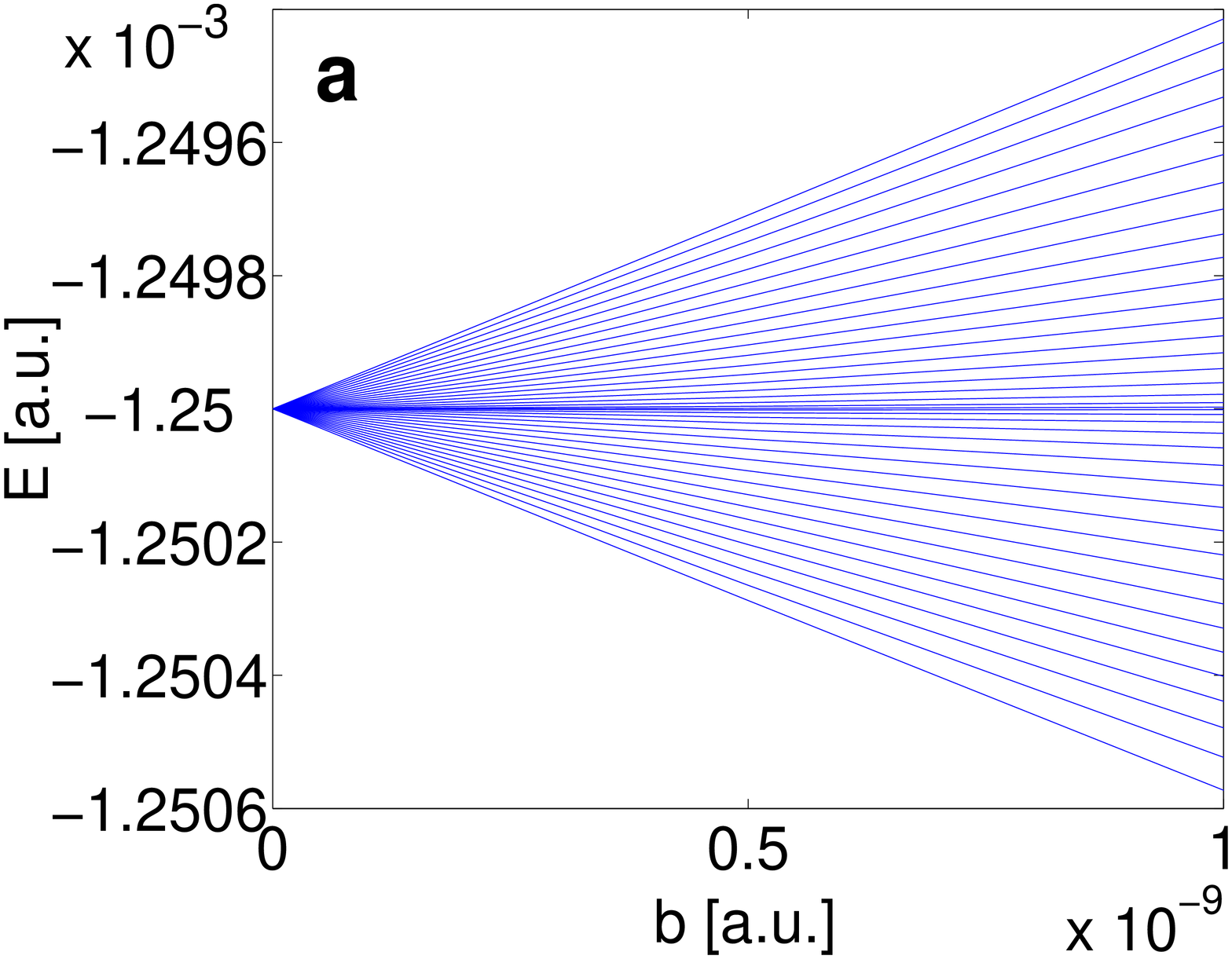}
\includegraphics[angle=-90,width=7.1cm]{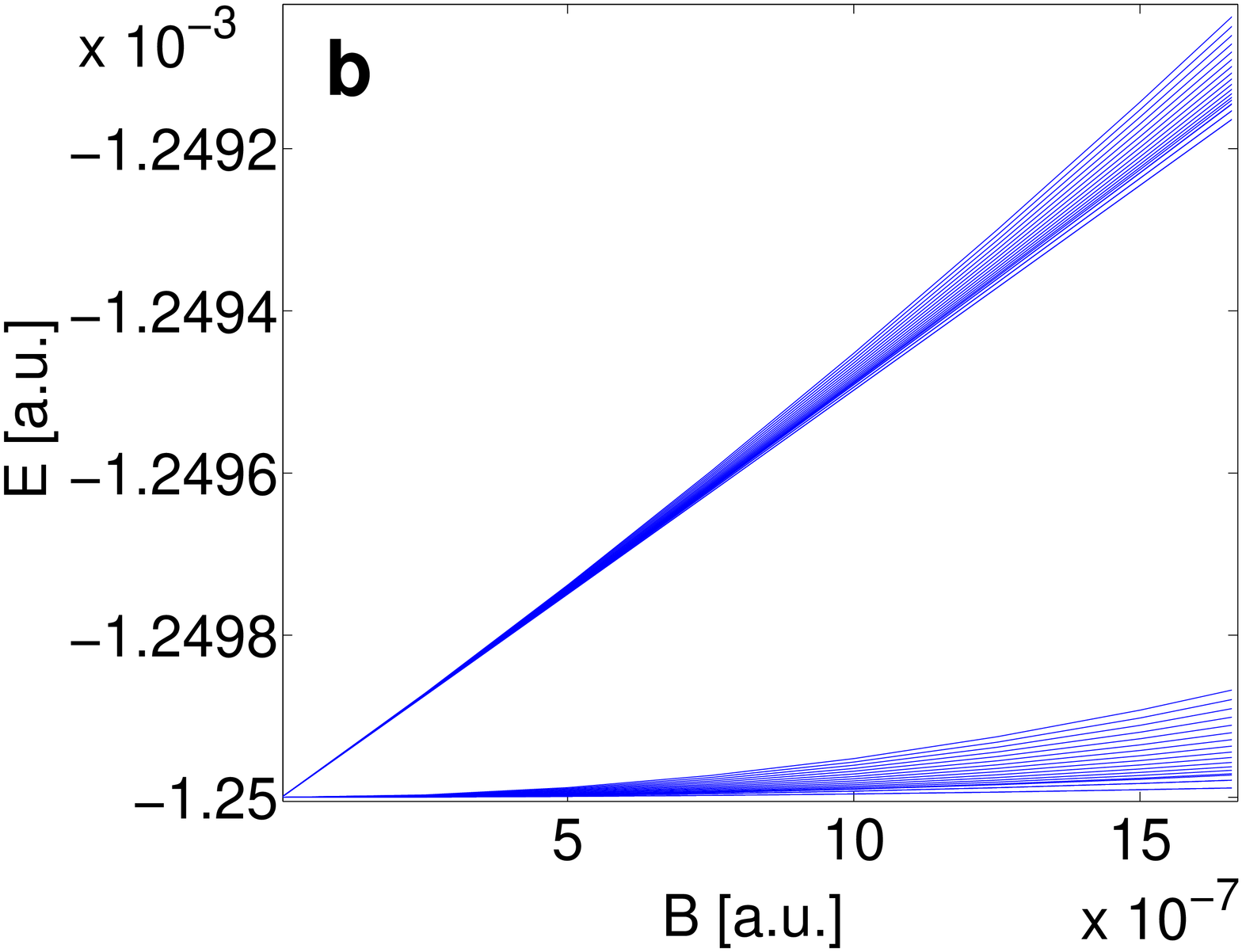}
\caption{\textbf{a}: Energy level splitting of the
$m=\frac{1}{2}$-states of the $n=20$-multiplet in the quadrupole
field. \textbf{b}: Splitting of the same energies in a homogeneous
field.}\label{fig:splitting_n20_b_and_B}
\end{figure}
Figure \ref{fig:splitting_n20_b_and_B}a shows the linear splitting
of the $n=20$ multiplet for the states with quantum number
$m=\pm\frac{1}{2}$. The levels are arranged symmetrically around
the energy for zero field. This is in contrast to the level
splitting of an atom in a homogeneous magnetic field which is
presented in figure \ref{fig:splitting_n20_b_and_B}b for the same
$n$-multiplet. Here the Zeeman-terms cause a linear energetical
splitting into only two branches. The latter (for fixed $m$) is
the result of the two possible spin orientations in a homogeneous
magnetic field. In a quadrupole field $L_z$ and $S_z$ are not
conserved separately and an assignment of spin orientations is
therefore not possible.

With increasing gradient the diamagnetic term with its quadratic
dependence on $b$ (or $B$ in the homogeneous field) becomes
increasingly important thereby resulting in a non-linear behavior
of the energy curves. The intra $n$-manifold mixing regime where
different $n$-multiplets are still energetically well separated
but different angular momentum states ($l$-states) mix can be
shown to scale as $b\propto n^{-6}$. In the high gradient regime
different $n$-multiplets begin to overlap and avoided crossings
dominate the spectra.
\begin{figure}[htbp]\center
\includegraphics[angle=0,width=16cm]{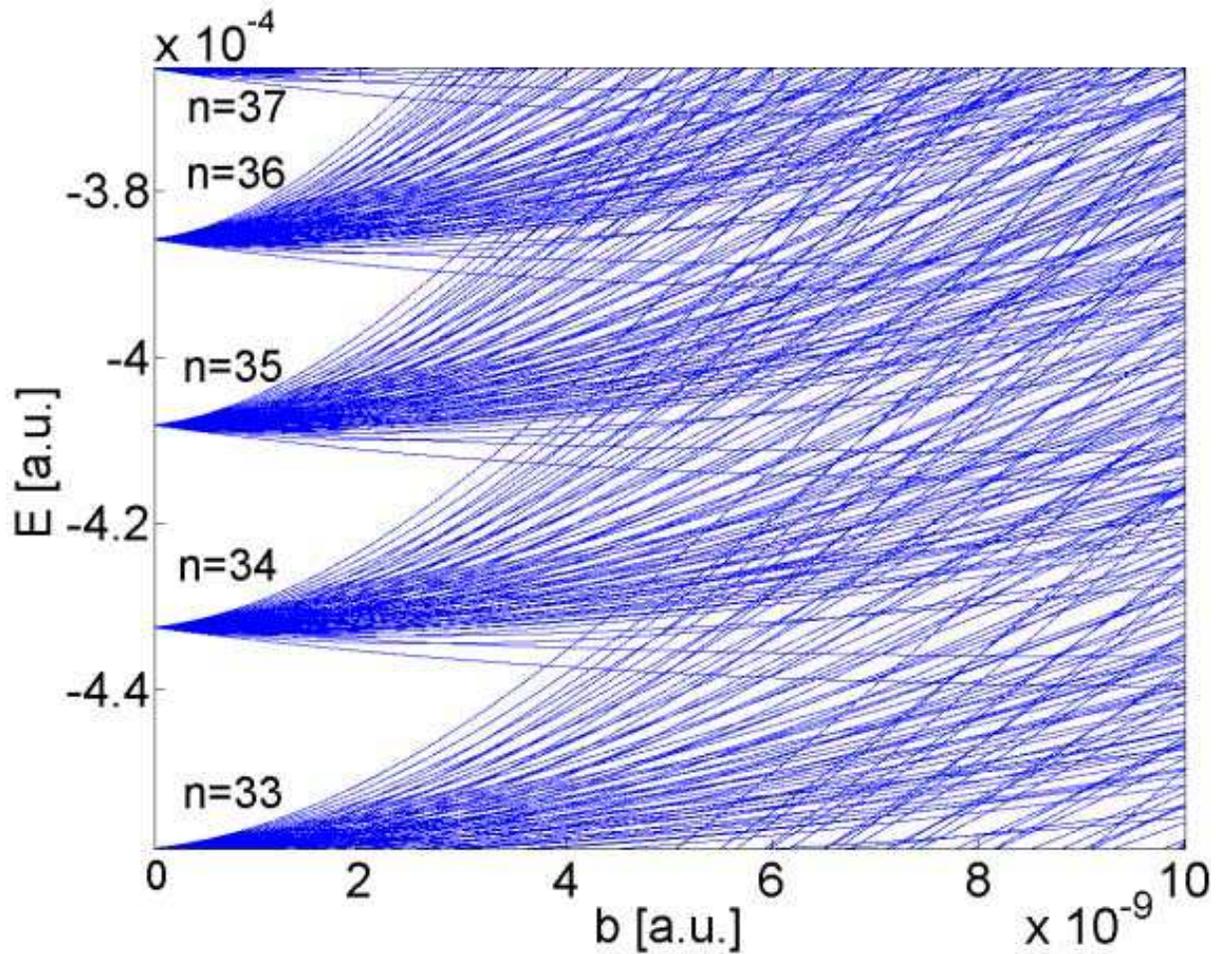}
\caption{Energy spectrum of the multiplets $n=33-37$ for
the states with $m=\pm\frac{1}{2}$. For low gradients the energy
levels split almost linearly followed by a transition region where
the diamagnetic term becomes increasingly more important. For high
gradients different $n$-multiplets overlap and no symmetries, i.e.
approximate quantum numbers, are left.}\label{fig:levels_n33-37}
\end{figure}
The onset of this inter $n$-manifold mixing scales according to
$b\propto n^{-\frac{11}{2}}$ whereas in the homogeneous field the
scaling is $B\propto n^{-\frac{7}{2}}$. Figure
\ref{fig:levels_n33-37} shows the energy levels of $n$-multiplets
ranging from $n=33$ to $n=37$ with $m=\pm\frac{1}{2}$ in the
regime $0 \le b \le 10^{-8}$. For such a degree of excitation the
intra $n$-manifold mixing sets in at $b\approx 5\cdot 10^{-10}$.
Therefore the linear splitting due to the Zeeman-term as shown in
figure \ref{fig:splitting_n20_b_and_B} is hardly visible.
%
%
\subsection{Low lying states and large gradients}\label{subsec:low_lying_states}
%
%
Although gradients $b\gg 10^{-10}$ are not experimentally
accessible it is at this point instructive to illuminate their
impact on low-lying electronic states. This way some principle
features of the influence of the quadrupole field on the
electronic structure of the atom will become clear.

\begin{figure}[ht]\center
\includegraphics[angle=0,width=7.1cm]{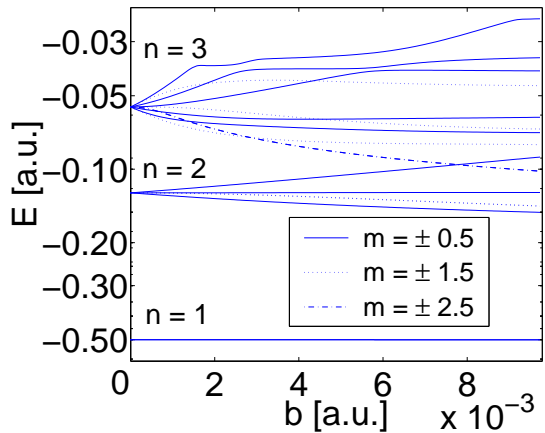}
\caption{Energy level structure of the three lowest
$n$-multiplets. The $J_z$ quantum number ranges from
$-\frac{5}{2}$ to $\frac{5}{2}$.}\label{fig:low_states}
\end{figure}

Here we consider the behavior of the lowest three $n$-multiplets
as a function of $b$ for $0<b<10^{-2}$ (see figure
\ref{fig:low_states}).
The splitting of the $n=3$ multiplet is linear up to $b=5\cdot
10^{-4}$, where inter $n$-manifold mixing begins.

The kink of the energetically highest-lying curve at $b=2\cdot
10^{-3}$ is the result of an avoided crossing with an energy level
from the upper $n=4$ multiplet. Moving to higher gradients we
encounter further avoided crossing. Crossings between levels
belonging to different quantum numbers $m$ are exact, since the
corresponding states belong to separate symmetry sub-spaces. The
$n=2$ multiplet does not show any indication of inter $n$-manifold
mixing throughout the whole gradient range considered here. Figure
\ref{fig:low_states} shows also that the energy of the degenerate
ground state depends very weakly on the gradient $b$: We observe a
maximum relative energy shift of about $0.04 \%$ at $b=10^{-2}$.
%
%
\subsection{Radial compression of the wavefunctions}\label{subsec:radial_compression}
\begin{figure}[htbp]\center
\includegraphics[angle=-90,width=7cm]{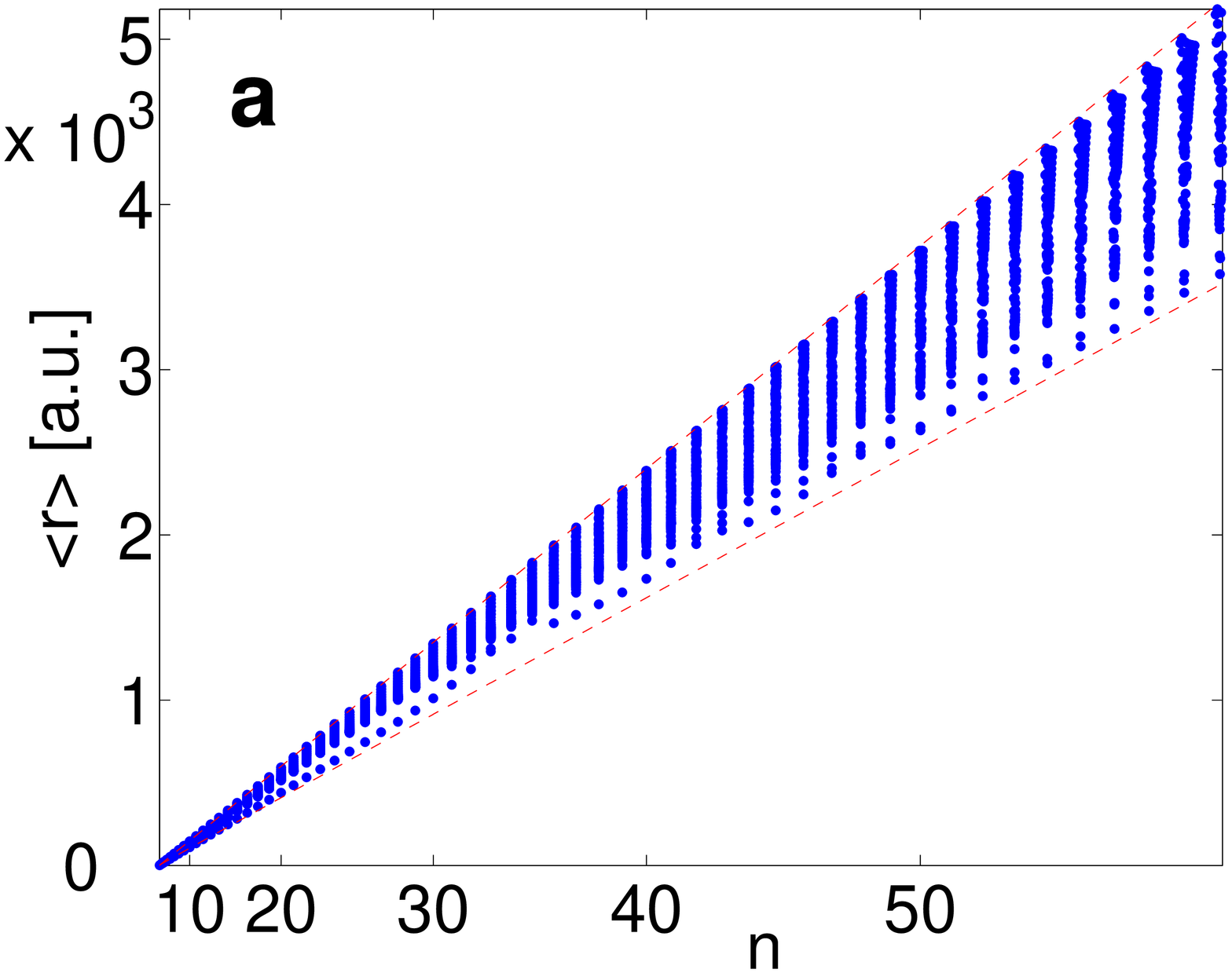}
\includegraphics[angle=-90,width=7cm]{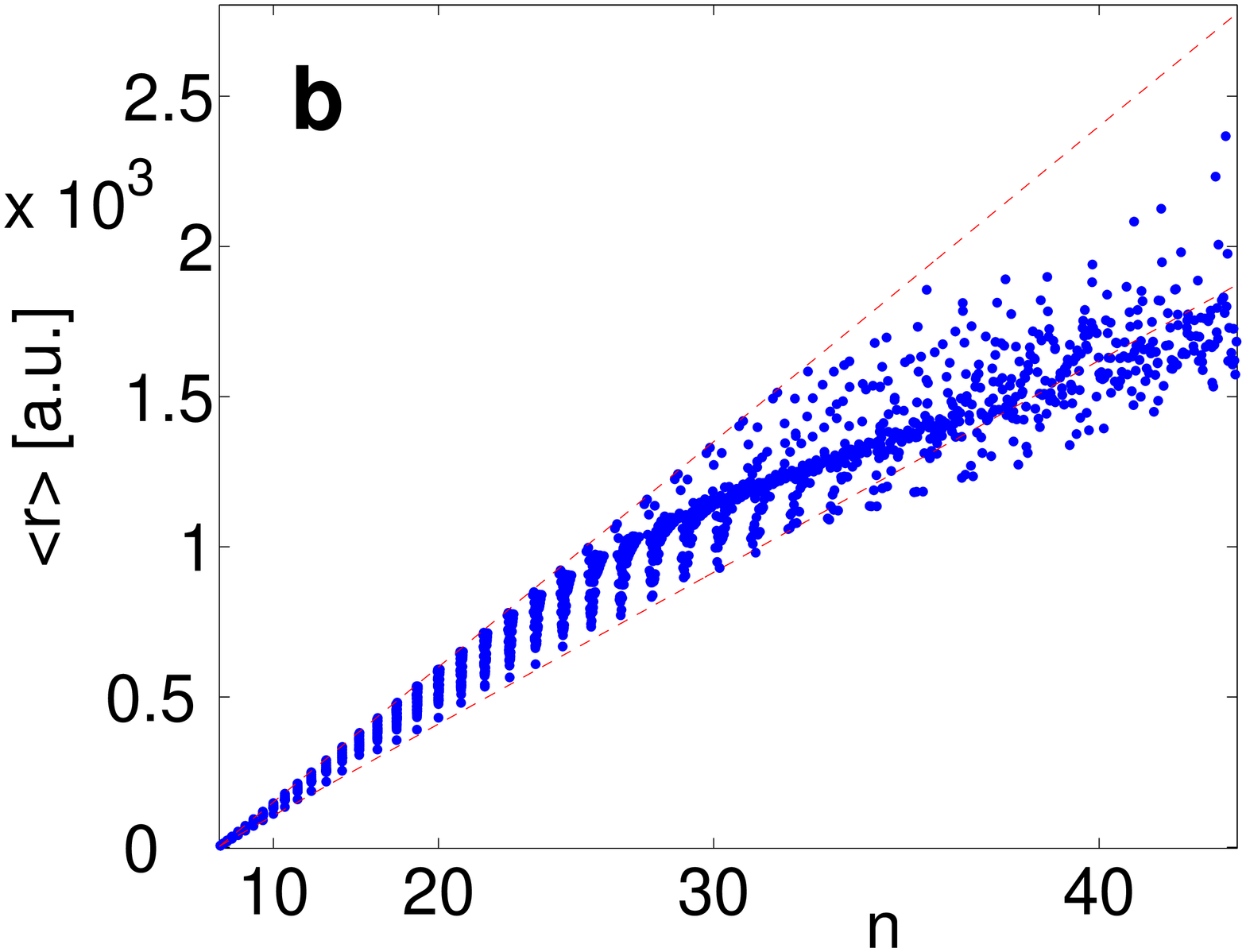}
\caption{Expectation value of the radial coordinate $r$ plotted
against the principle quantum number $n$ at different gradients
(\textbf{a:} $b=10^{-10}$, \textbf{b:}
$b=10^{-8}$).}\label{fig:radial_expectation_value}
\end{figure}

%
%
The operator $r$ obeys the commutators $\left[r,TOP_z\right]=0$
and $\left[r,J_z\right]=0$.
Together with (\ref{eq:observable_property_2}) this yields
\begin{eqnarray}
\left<r\right>=\left<r\right>^\pm_{TOP_z}=\left<r\right>_{J_z}.
\end{eqnarray}
Thus the expectation value of $r$ in the $J_z$ eigenstates is the
same as in the $TOP_z$-eigenstates.\\
For the one-electron problem without external field we have
$\left<r\right>^H=\frac{1}{2}\left(3n^2-l(l+1)\right)$. Since $l$
satisfies $0\leq l \leq n-1$ one obtains corresponding upper and
lower bounds for the expectation value of $r$:
$n\left(n+\frac{1}{2}\right)\leq \left<r\right>^H \leq
\frac{3}{2}n^2$. In figure \ref{fig:radial_expectation_value} this
expectation value is shown for the atom in the quadrupole field as
a function of the principle quantum number $n$ for two different
gradients ($n$ serves here as a label for the energy levels and is
not a good quantum number!).
The radial expectation values lie in between the boundaries given
by the field-free inequality indicated by the dashed lines (see figure
\ref{fig:radial_expectation_value}a).
Expectation values of states belonging to the same $n$-multiplet
are arranged in vertical lines expressing their energetical
degeneracy (see figure \ref{fig:radial_expectation_value}a for
$b=10^{-10}$). Here states with large expectation values of $r$
possess small expectation values of the angular momentum and vice
versa. The situation is different for $b=10^{-8}$. Here a
systematic decrease of the $r$-expectation values takes place
for states with an energy corresponding to $n> 30$. This is also
the energy regime where the inter $n$-manifold mixing sets in.
Here the mentioned pattern of vertical lines dissolves and a irregular
looking distribution emerges. Although the expectation values
are not located between the dashed boundaries the distribution is
still widespread which can be ascribed to the different angular
momentum distribution of the states.

We now investigate the deformation of the ground state
wavefunction in both the quadrupole and the homogeneous field. Due
to their simple nodal structure they are suited best to point out
differences arising for the two different field configurations.
To demonstrate the deformation effects on the electronic cloud
we consider for reasons of illustration an extremely high gradient.
\begin{figure}[htbp]\center
\includegraphics[angle=0,width=7cm]{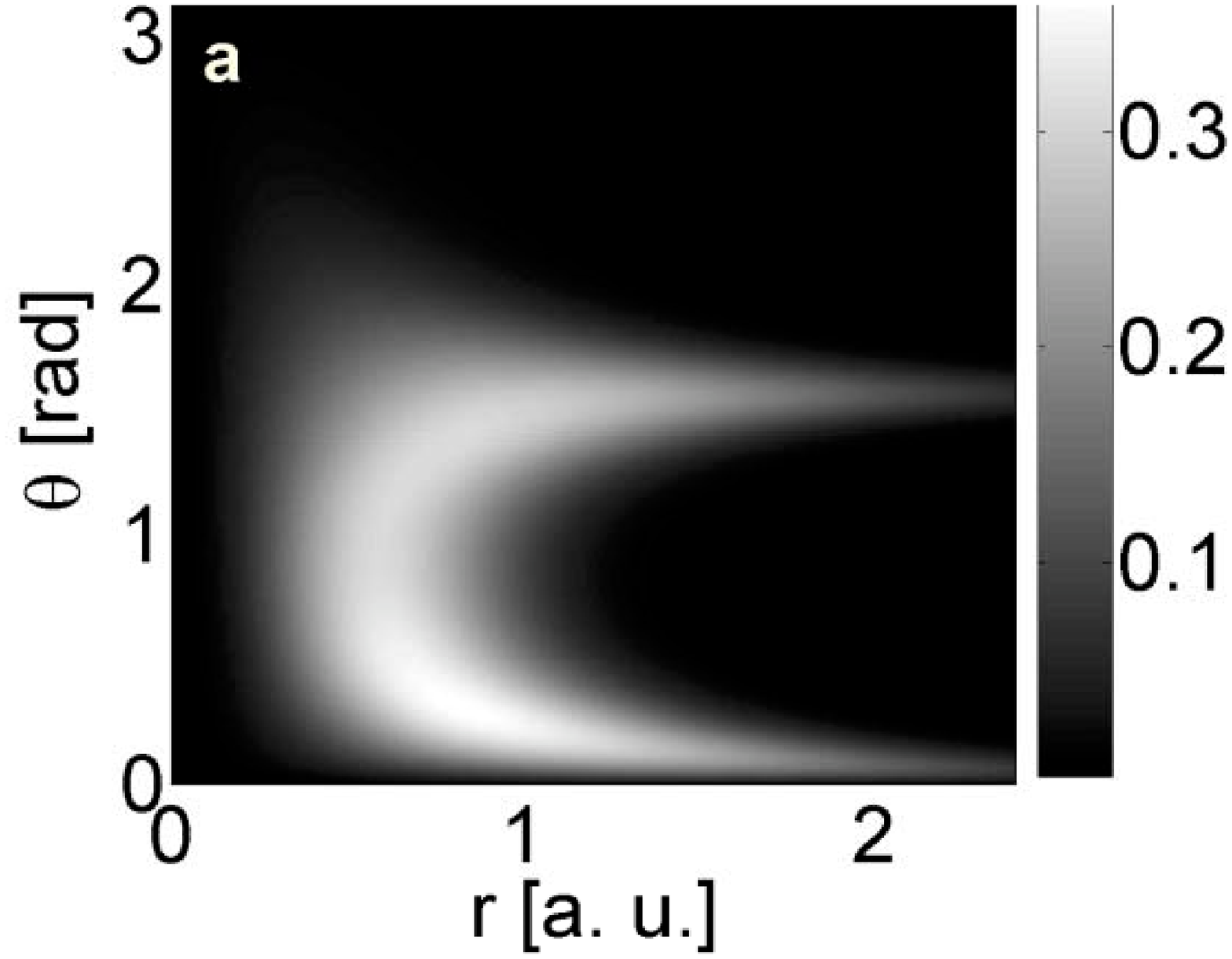}
\includegraphics[angle=0,width=7cm]{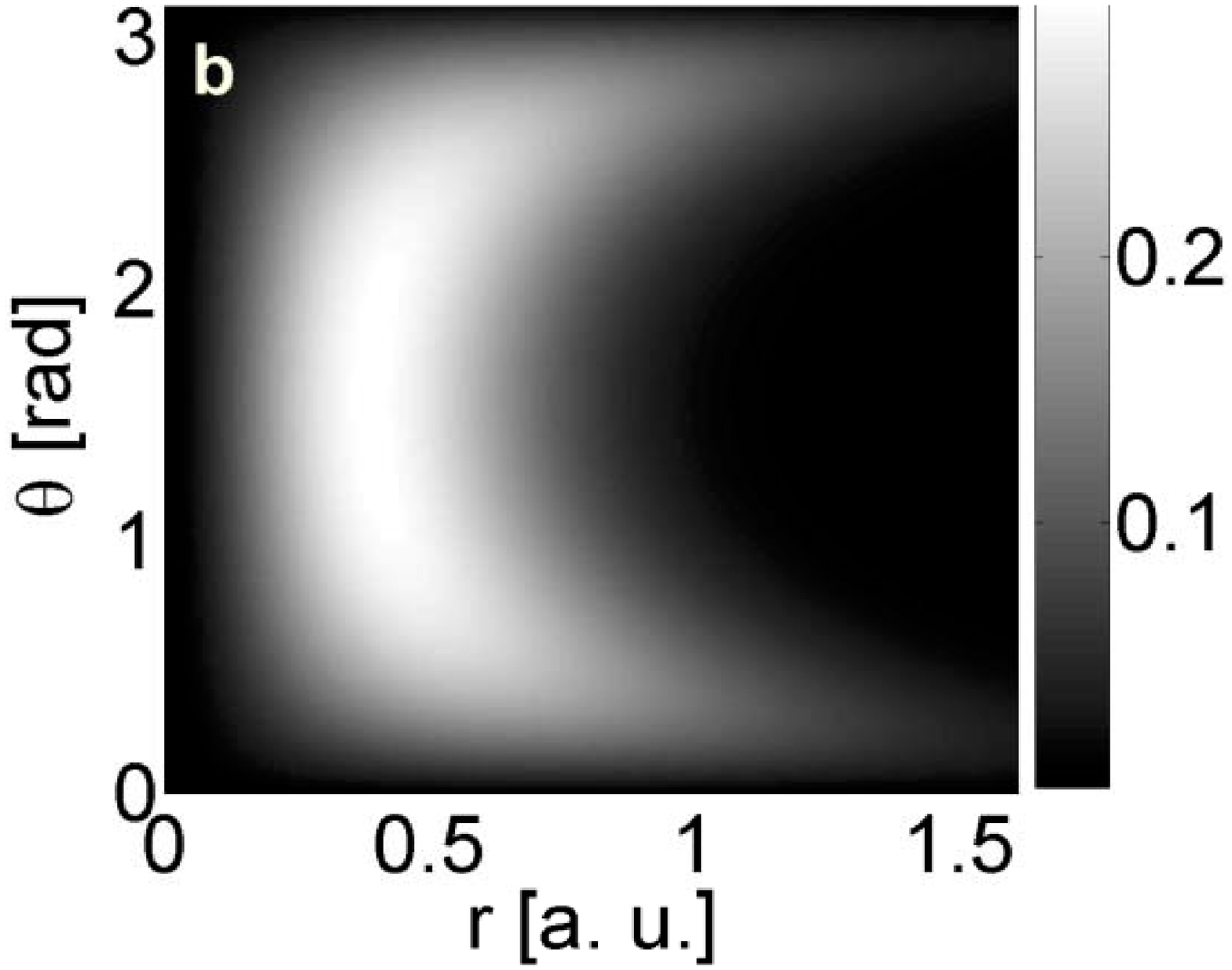}
\caption{Spatial probability density of the ground state in the
quadrupole field (\textbf{a}) and a homogeneous field pointing in
the $z$-direction (\textbf{b}) for $b=B=10$ and
$m=\frac{1}{2}$.}\label{fig:groundstate}
\end{figure}
Figure \ref{fig:groundstate} shows the spatial probability density
of the ground state in the quadrupole field
(\ref{fig:groundstate}a) at $b=10$ and the homogeneous field
(\ref{fig:groundstate}b) at $B=10$ ($m=\frac{1}{2}$ in both
cases).

For the quadrupole field we observe an asymmetric deformation with
respect to the $\theta=\frac{\pi}{2}$-plane: the electronic
wavefunction is almost completely confined to the upper
half-volume ($\theta<\frac{\pi}{2}$) which is a consequence of the
symmetries discussed in section \ref{sec:symmetries}: $z$-parity
is not conserved and eigenstates appear in pairs one being the
mirror image of the other with respect to the reflection at the
$x-y$-plane. Furthermore we observe that the electronic motion is
localized particularly along the two 'channels' for $\theta=0$
corresponding to the lower $z$-axis and $\theta=\frac{\pi}{2}$
being the $x-y$-plane. This property as well as the detailed shape
of the electronic probability density in the individual
half-volume is determined by the diamagnetic term which is
dominant in the high gradient regime. For the quadrupole field it
is proportional to $\sin^2\theta\cos^2\theta$ reaching its maximum
value at $\theta=\frac{\pi}{4},\frac{3\pi}{4}$. The probability
density in a homogeneous field (see figure \ref{fig:groundstate}b) exhibits the
above-mentioned corresponding reflection symmetry due to the
invariance of the corresponding Hamiltonian with respect to
$z$-parity. Here we observe the maximum of the probability density
at $\theta=\frac{\pi}{2}$ and a deformation towards $\theta=\pi$
and $\theta=0$ leading to a cigar-like shape. The diamagnetic term
is proportional to $\sin^2\theta$ having its maximum value at
$\theta=\frac{\pi}{2}$ thus coinciding with the regions possessing
the strongest deformation of the probability density. For both
field configurations the probability density vanishes at $r=0$.
The maximum is reached at $r\approx 0.75$ in the quadrupole field
and for $r\approx 0.5$ in the homogeneous field followed by an
exponential drop-off.

%
%
\subsection{Properties of the electronic spin}\label{subsec:spin_properties}
%
%
%
%
\subsubsection{Expectation values}\label{subsubsec:spin_expectation_value}
%
%
In a homogeneous magnetic field the projection of the spin
operator onto the direction of the magnetic field
is a conserved quantity. In this case one can choose the energy
eigenstates to be also eigenfunctions with respect to $S_z$. This
restricts $\left<S_z\right>$ to the two values $\pm\frac{1}{2}$.
In the quadrupole field $S_z$ is not conserved, and we therefore consider
the expectation value $\left<S_z\right>$ of the
electronic states. We have
\begin{eqnarray}
\left<S_z\right>_{J_z}=\frac{1}{2}\left[\left<u\mid
u\right>-\left<d\mid d\right>\right].
\end{eqnarray}
\begin{figure}[ht]\center
\includegraphics[angle=0,width=7cm]{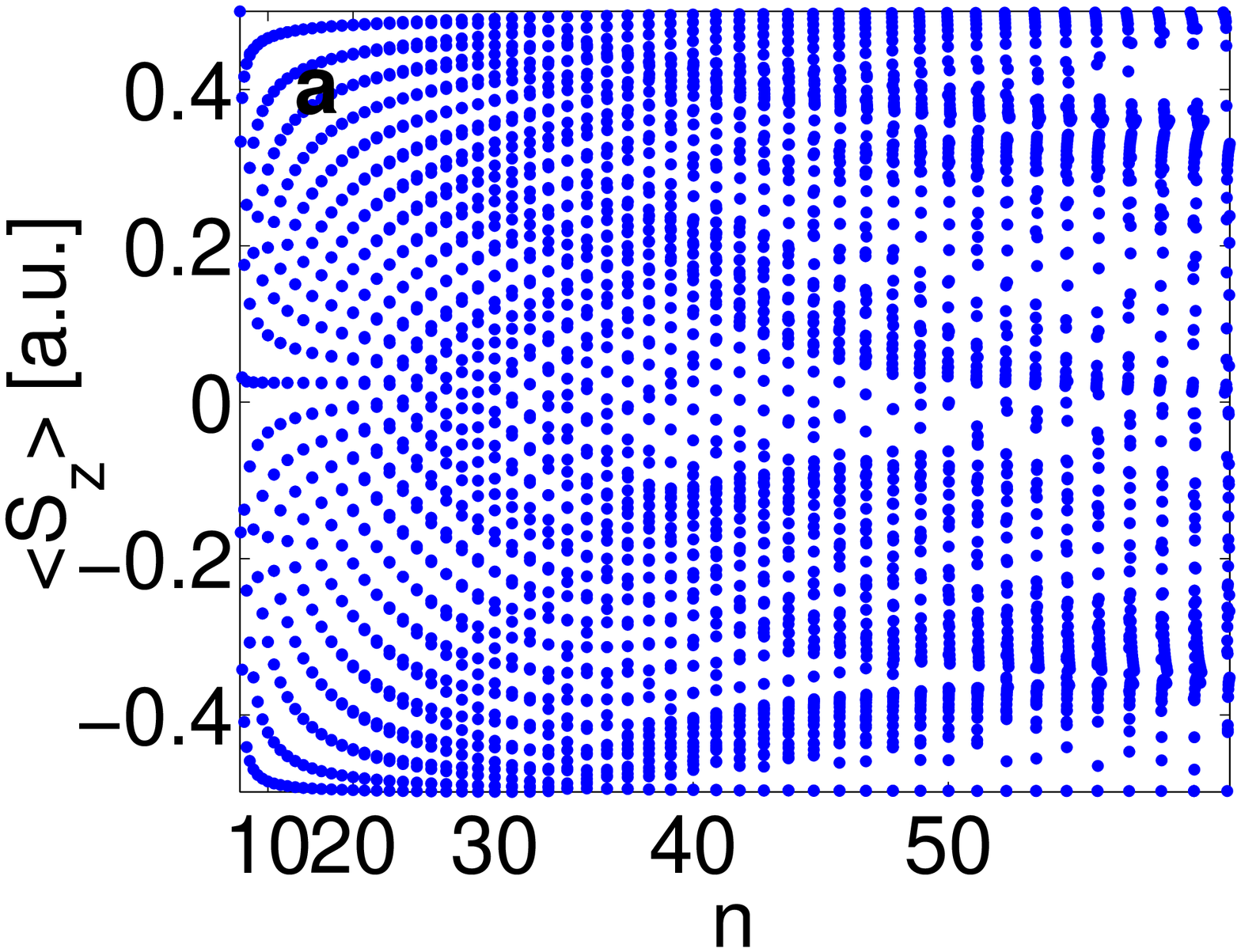}
\includegraphics[angle=0,width=7cm]{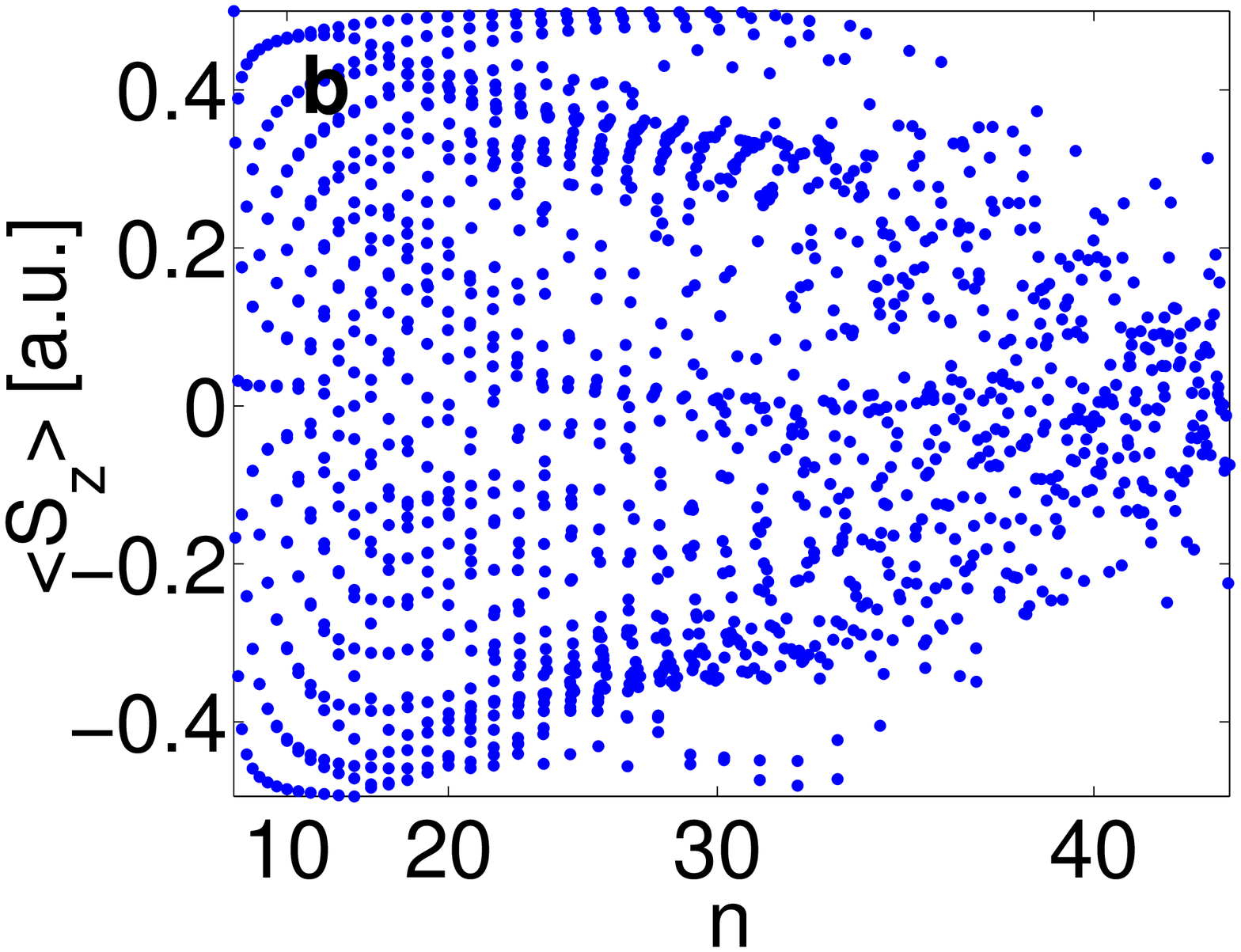}
\caption{Expectation values of the $z$-component of the spin
operator as a function of the quantum number $n$ for different
gradients (\textbf{a}: $b=10^{-10}$, \textbf{b}: $b=10^{-8}$).
}\label{fig:sz_expectation_value}
\end{figure}
Figure \ref{fig:sz_expectation_value} shows the distribution of
$\left<S_z\right>_{J_z}$ for electronic states of the $m=\frac{1}{2}$
subspace as a function of the principle quantum number $n$. Since
$S_z$ is not conserved the values of $\left<S_z\right>$ are
allowed to cover the complete interval
$\left[-\frac{1}{2},\frac{1}{2}\right]$. For
$b=10^{-10}$ (figure \ref{fig:sz_expectation_value}a) and a low
degree of excitation the expectation values are evenly distributed
over the interval. When reaching highly excited states this
pattern becomes increasingly distorted. The expectation values
agglomerate at $-0.35$ and $0.35$ for $n\geq 50$. Due to the
approximate degeneracy of the energy levels at low energies, i.e.
small $n$, the values of $\left<S_z\right>$ form vertical lines.
These lines widen for higher $n$. Since at $b=10^{-10}$ no
significant $n$-mixing up to our maximum converged energy levels
takes place, neighboring lines are well separated.
For a higher gradient $b=10^{-8}$ (figure
\ref{fig:sz_expectation_value}b) the above properties are equally
present for low-lying states. However, with increasing
excitation energy we now observe a complete $n$-mixing  regime
where the regular patterns disappear and we obtain an irregular
distribution of $\left<S_z\right>$. Overall the distribution
narrows, e.g. for $n=40$ the occupied interval is approximately
$\left[-0.3,0.3\right]$.

We remark that due to the fact that the $S_z$-operator
anti-commutes with $TOP_z$ we have
$\left<S_z\right>_{TOP_z}^\pm=0$. Apparently there is no preferred
direction for the electronic spin in a state obeying the $TOP_z$
symmetry.

%
%
\subsubsection{Spin polarization}\label{subsubsec:spin_sz_density}
%
%
Due to the coupling of the spatial to the spin degrees of freedom
we have to expect a dependence of the spin orientation on the
spatial coordinates. To investigate this in more detail we study
the spatially dependent $S_z$-polarization $W_S(\vec{r})$. For a $J_z$-eigenstate
$\left|E,m\right>$ it reads
\begin{eqnarray}
W_S(\vec{r})&=&\frac{\left<E,m\mid \vec{r}
\right>{S}_z\left<\vec{r}\mid E,m\,\right>}{\left<E,m\mid
\vec{r}\right>\left<\vec{r}\mid E,m\,\right>}
=\frac{1}{2}\frac{\left|\left<u\mid
\vec{r}\right>\right|^2-\left|\left<d\mid
\vec{r}\right>\right|^2}{\left|\left<u\mid
\vec{r}\right>\right|^2+\left|\left<d\mid
\vec{r}\right>\right|^2}.\label{eq:spin_density}
\end{eqnarray}
\begin{figure}[ht]\center
\includegraphics[angle=0,width=7cm]{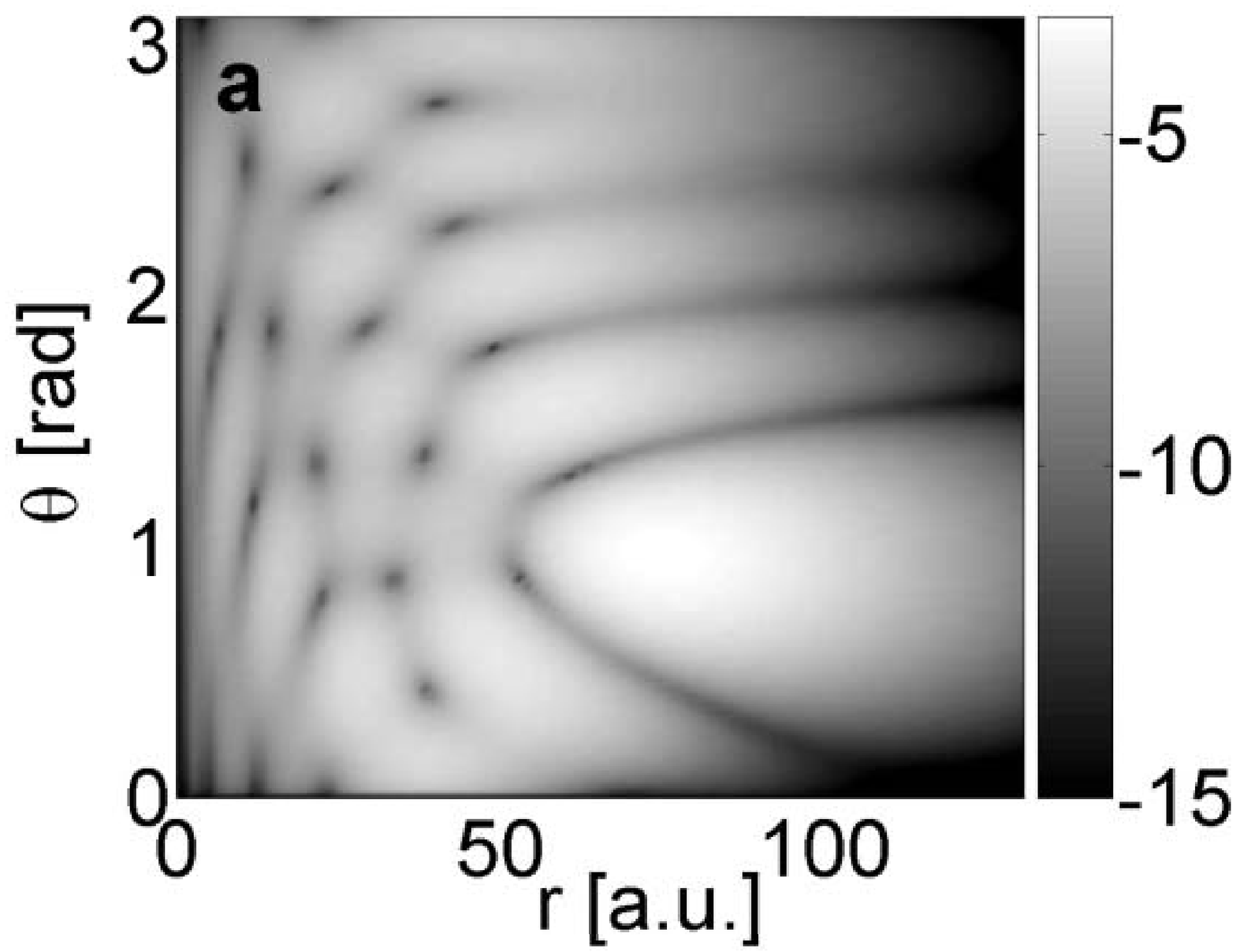}
\includegraphics[angle=0,width=7cm]{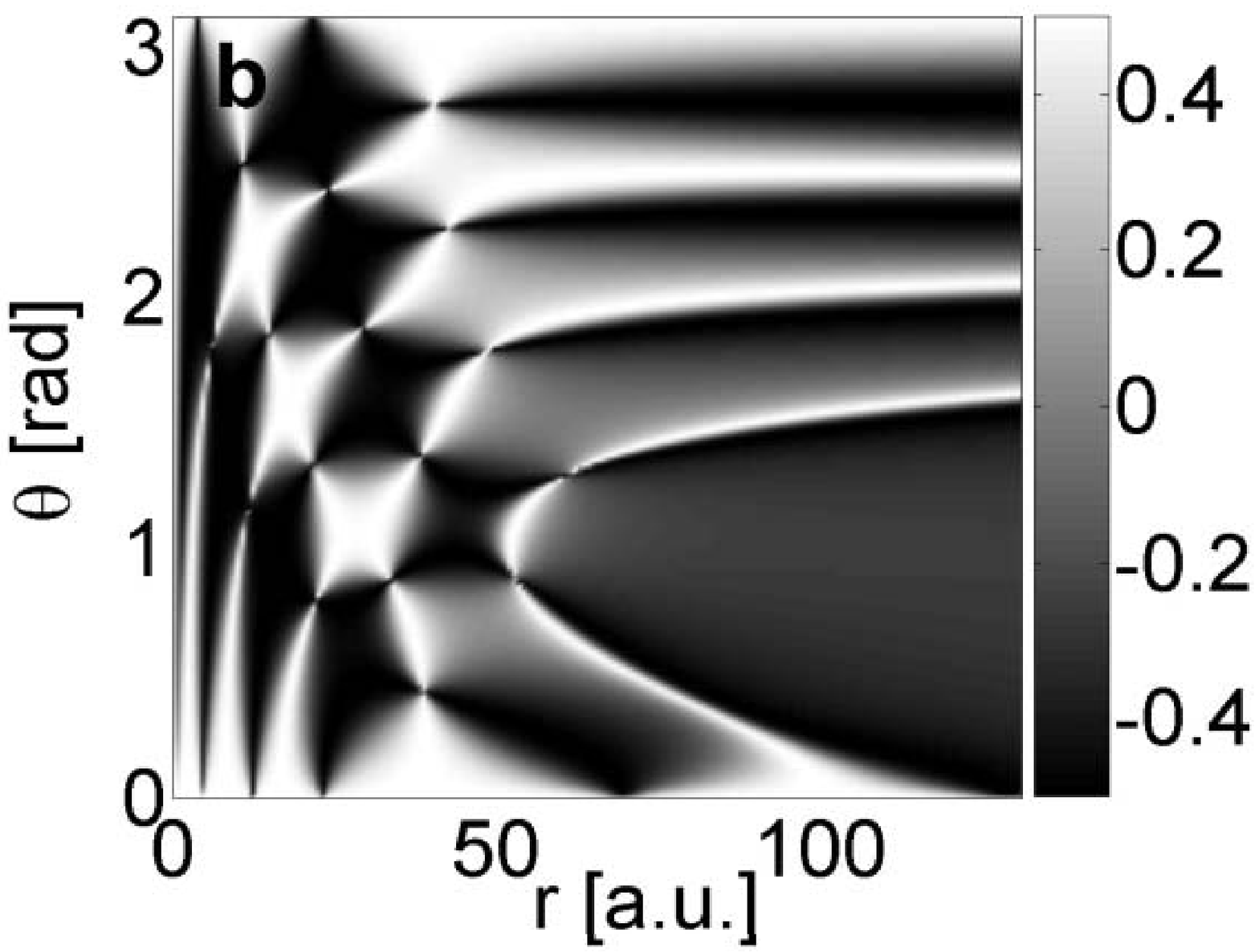}
\caption{\textbf{a}: Spatial probability density
of the $45th$ excited state for $m=\frac{1}{2}$ and
$b=10^{-8}$ in logarithmic representation. \textbf{b}:
$S_z$-polarization for the same state.}\label{fig:sz_density_1}
\end{figure}
Figure \ref{fig:sz_density_1} shows the spatial probability
distribution (\ref{fig:sz_density_1}a) and the $S_z$-polarization
(\ref{fig:sz_density_1}b) for the $45th$ excited state for
$m=\frac{1}{2}$ and $b=10^{-8}$ emerging from the $n=7$ multiplet
at $b=0$. The probability density is shown for a logarithmic scale
to provide details for small values. In contrast to the constant
$S_z$-polarization we would encounter in the absence of a field or
a homogeneous field we observe a complex pattern of domains
exhibiting different spin orientation (white: spin up, black: spin
down). At low $r$ values these domains form a pattern similar to
that of a chess board. With increasing $r$ a transition region is
encountered where the formation of stripes with different spin
orientation begins. The junctions where four spin domains meet
each other coincide with the nodes of the spatial probability
density. The Coulomb interaction as well as the spin-Zeeman term
are responsible for the interwoven network of island of different
spin orientation. The additional presence of the orbital Zeeman
and the diamagnetic term leads to a deformation of this network.

\begin{figure}[ht]\center
\includegraphics[angle=0,width=7cm]{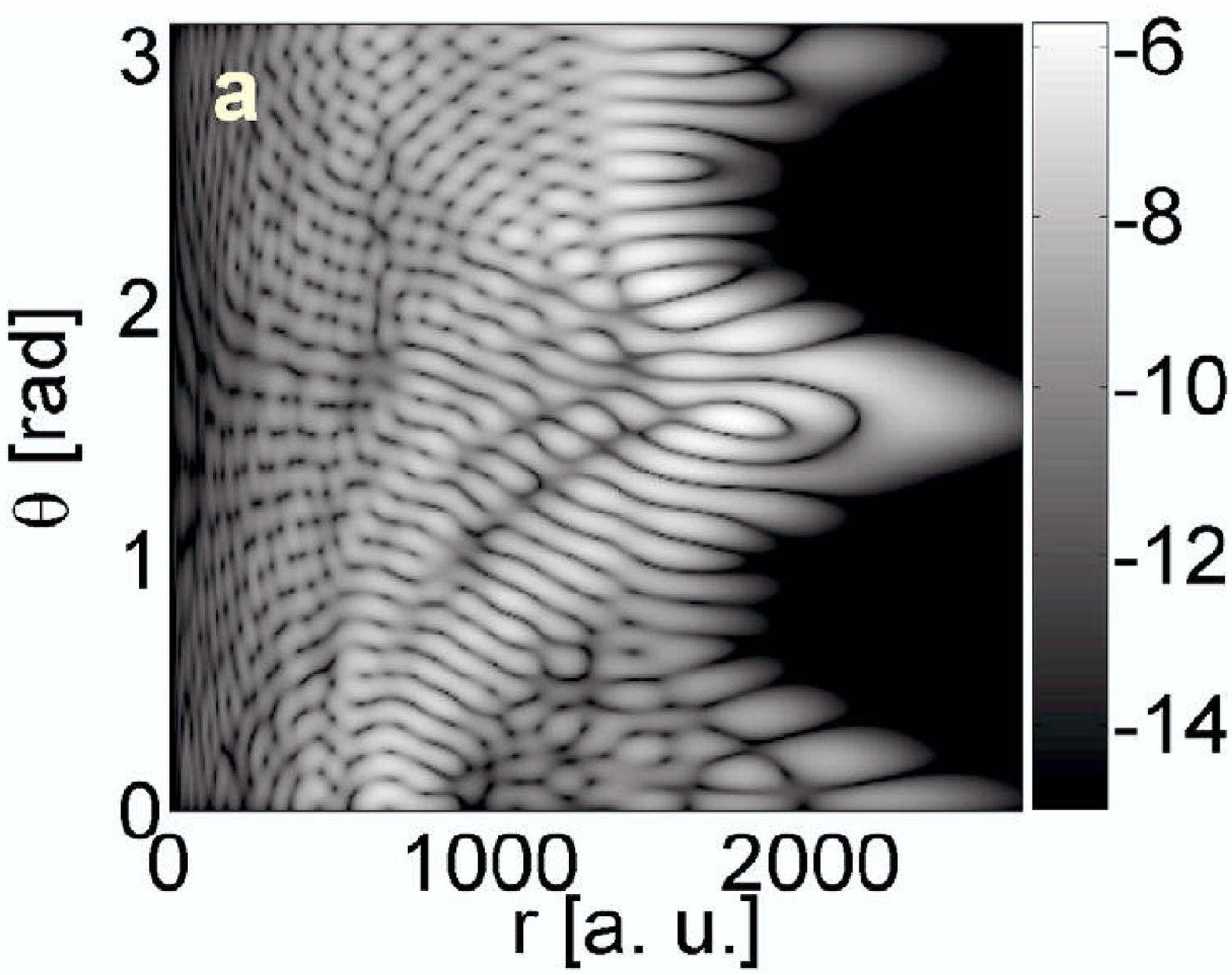}
\includegraphics[angle=0,width=7cm]{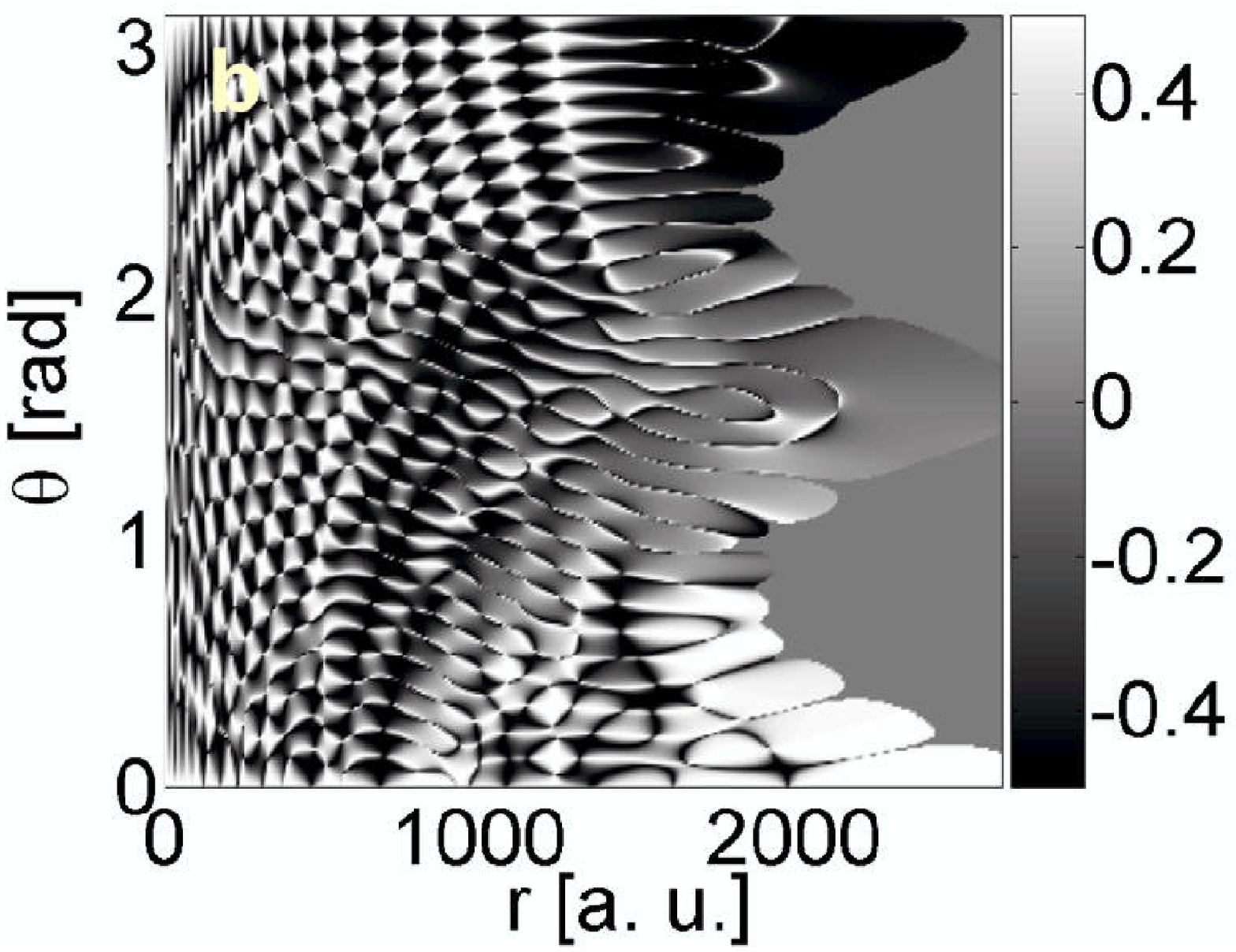}
\caption{Spatial probability density (\textbf{a}) and
$S_z$-polarization (\textbf{b}) for the $1117th$ excited state for
$m=\frac{1}{2}$ and $b=10^{-8}$. At large $r$ the $S_z$-polarization
becomes similar to $W_S^+$ indicating an antiparallel alignment of
the electronic spin to the magnetic
field.}\label{fig:sz_density_2}
\end{figure}

Highly excited states, i.e. Rydberg states, typically possess
a large spatial extension. In
figure \ref{fig:sz_density_2} the spatial probability density
(\ref{fig:sz_density_2}a) and the $S_z$-polarization
(\ref{fig:sz_density_2}b) are depicted for the $1117th$ excited
state of the $m=\frac{1}{2}$ subspace ($b=10^{-8}$). At small $r$-values
the $S_z$-polarization shows the chess board pattern, we have already
observed for the low-lying excited states, followed by a striped region.
Whereas for low radii the spin orientation changes locally from
island to island generating an appealing pattern we observe an overall tendency
of the electronic spin polarization in the region characterized by stripes ($r \approx 2000$).
Here, independently of the nodal structure, the spin orientation
changes smoothly from downwards at $\theta=\pi$ to upwards at $\theta=0$.
This feature can be understood by inspecting the spin Zeeman term only.
The corresponding Hamiltonian reads
\begin{eqnarray}
\hat{H}_S&=&-\vec{\mu}\,\vec{B}(\vec{r})=\frac{b}{2}r\left(%
\begin{array}{cc}
  -2\,\cos\theta & \sin\theta\,e^{-i\phi} \\
  \sin\theta\,e^{i\phi} & 2\,\cos\theta \\
\end{array}%
\right)\label{eq:mu_B_coupling_Hamiltonian}.
\end{eqnarray}
Here the complete dynamics takes place in spin space since the
spatial coordinates $\left(r,\theta,\phi\right)$ are entering
as parameters. The eigenvalue problem belonging to
(\ref{eq:mu_B_coupling_Hamiltonian}) possesses the two solutions
\begin{eqnarray}
\Phi_\pm(r,\theta,\phi)=\left(%
\begin{array}{c}
  -\frac{2\,\cos\theta\pm\sqrt{1+3\,\cos^2\theta}}{\sin\theta}e^{-i\phi} \\
  1 \\
\end{array}%
\right)
\end{eqnarray}
with the energies
\begin{eqnarray}
E_\pm=\mp\frac{1}{2}b\,r\sqrt{1+3\cos^2\theta}=\mp|\vec{\mu}||\vec{B}(\vec{r})|.\label{eq:approx_eigenstates}
\end{eqnarray}
These energies correspond to those of a spin oriented parallel
($E_-$) or antiparallel ($E_+$) to the magnetic quadrupole field.
Constructing the $S_z$-polarization $W_S^\pm$ of the eigenstates
(\ref{eq:approx_eigenstates}) yields
\begin{eqnarray}
W_S^\pm=\frac{1}{2}\frac{(2\cos\theta\pm\sqrt{1+3\,\cos^2\theta})^2-\sin^2\theta}{(2\cos\theta\pm\sqrt{1+3\,\cos^2\theta})^2+\sin^2\theta}.\label{eq:spindichte_naeherung}
\end{eqnarray}
\begin{figure}[ht]\center
\includegraphics[angle=0,width=7cm]{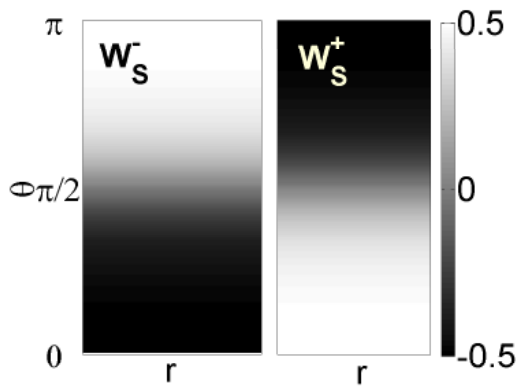}
\caption{$S_z$-polarization of the two eigenstates of the Hamiltonian
(\ref{eq:mu_B_coupling_Hamiltonian}). The electronic spin either
points parallel ($W_S^-$) or antiparallel ($W_S^+$) in the local
direction of the field.}\label{fig:sz_density_approx}
\end{figure}
Both $W_S^+$ and $W_S^-$ are shown in figure
\ref{fig:sz_density_approx}. They do not depend on the radial
coordinate and the azimuthal angle $\phi$. The $z$-component
of the spin is oriented upwards/downwards at $\theta=0$ and points
in the opposite direction at $\theta=\pi$. For
$\theta=\frac{\pi}{2}$ there is a transition region with $W_S^\pm$
being close to zero.

This is precisely the above-discussed behavior we observe at large
$r$ for the $S_z$-polarization of the Rydberg state depicted in
figure \ref{fig:sz_density_2}b. Apparently in this special state
the electron spin prefers an antiparallel alignment to the
external field at large radii which corresponds to $W_S^+$. We
remark that similar results on the spatial spin polarization are
obtained if one considers the quantity $\vec{S}\vec{B}(\vec{r})$
instead of $S_z$.

%
%
\subsection{Electromagnetic transitions}\label{subsec:em_transitions}
%
%

The interaction of an atom in the magnetic field with external
electromagnetic radiation leads to transitions among its
electronic states which we describe employing the dipole
approximation. The amplitude for a transition from the initial
state $\left|i\right>$ to the final state $\left|f\right>$ is
given by the matrix element
$p_{if}^{\sigma^\pm,\pi}=\left<i\right|D_{\sigma^\pm,\pi}\left|f\right>$.
Considering $\pi$ and $\sigma^\pm$ transitions the operator takes
the common forms $D_\pi=z=r\cos\theta$ and
$D_{\sigma^\pm}=\frac{1}{\sqrt{2}}\left(x\pm i
y\right)=\frac{1}{\sqrt{2}}r\sin\theta e^{\pm i\phi}$,
respectively. The  symmetries of the $J_z$ and $TOP_z$ eigenstates
result in selection rules which we will discuss in the following.

Evaluating the transition matrix element for $\pi$-transitions
$\left<E^\prime,m^\prime\right|r\cos\theta\left|E,m\right>$ leads
to the selection rule $m^\prime-m=0$. For $\sigma^\pm$-transitions
the corresponding matrix element reads
$\left<E^\prime,m^\prime\right|r\sin\theta\,e^{\pm
i\phi}\left|E,m\right>$ which is only non-zero if
$m^\prime-m=\pm1$. Hence, electromagnetic transition between $J_z$
states occur only if $\triangle m=0,\pm1$ reminiscent of the
situation in a homogeneous or even no magnetic field. The
corresponding matrix element for $\pi$-transitions between
$TOP_z$-eigenstates is
$\left<E^\prime,\Pi^\prime\right|r\cos\theta\left|E,\Pi\right>$
leading to the selection rule $\Pi^\prime\neq\Pi$, i.e. only
transitions between states with opposite $TOP_z$ symmetry are
allowed. There are no selection rules for $\sigma^\pm$-transitions
between $TOP_z$-eigenstates. The selection rules are summarized in
table \ref{tbl:selection_rules}.
\begin{table}[ht]\center
\begin{tabular}{|c||c||c|}
\hline & $J_z-states$ & $TOP_z-states$\\
\cline{2-3} \raisebox{2.3ex}[-2.3ex]{transition type}& $\triangle m=m'-m$ & $\triangle \Pi=\Pi' -\Pi$\\
\hline\hline   $\pi$  & 0 & $\pm 2$\\
\hline $\sigma^+$  & 1 & -\\
\hline $\sigma^-$  & -1 & -\\
\hline
\end{tabular}
\caption{Selection rules for dipole transitions between
$J_z$-eigenstates and between
$TOP_z$-eigenstates.}\label{tbl:selection_rules}
\end{table}

\begin{figure}[ht]\center
\includegraphics[angle=0,width=7.3cm]{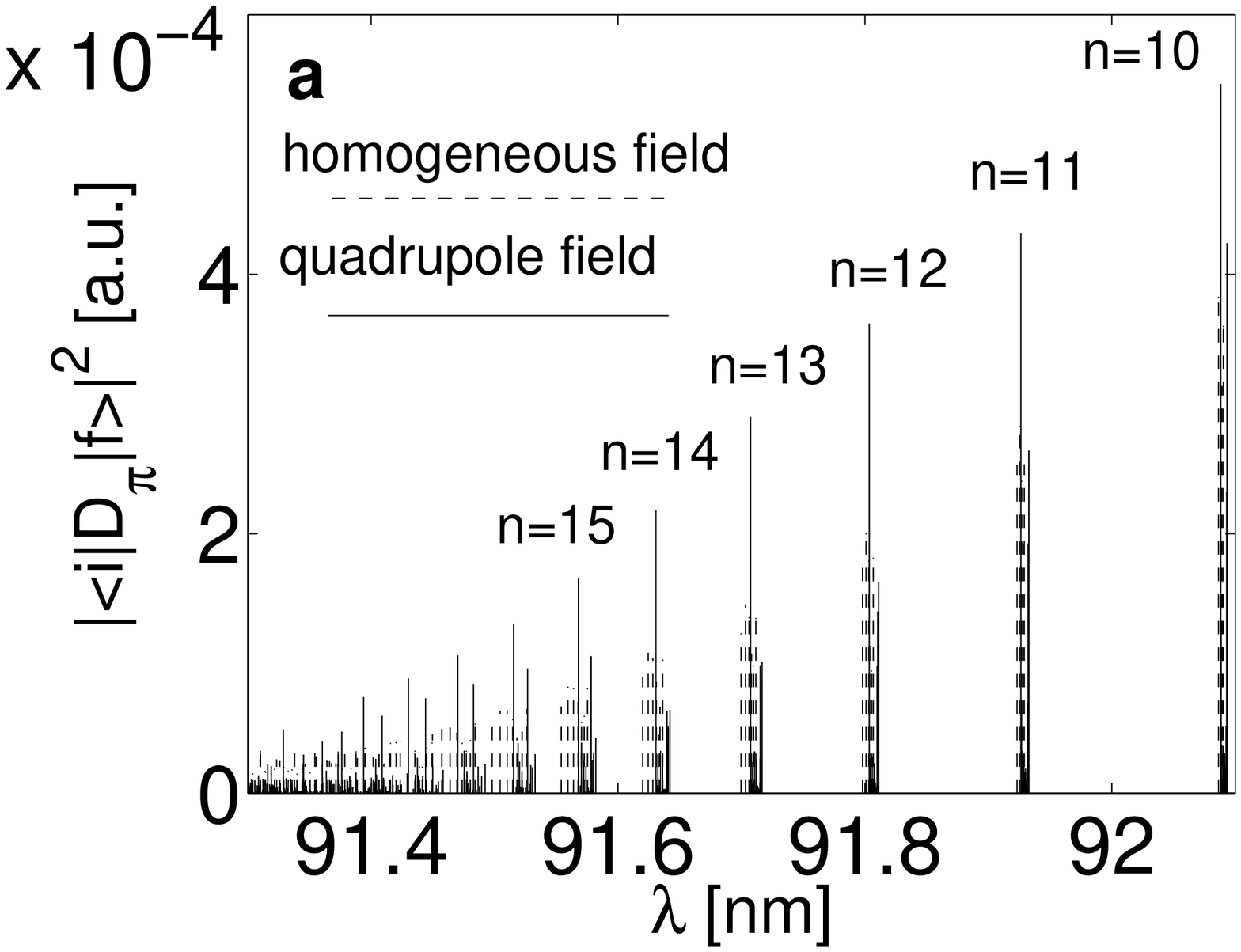}
\includegraphics[angle=0,width=7.3cm]{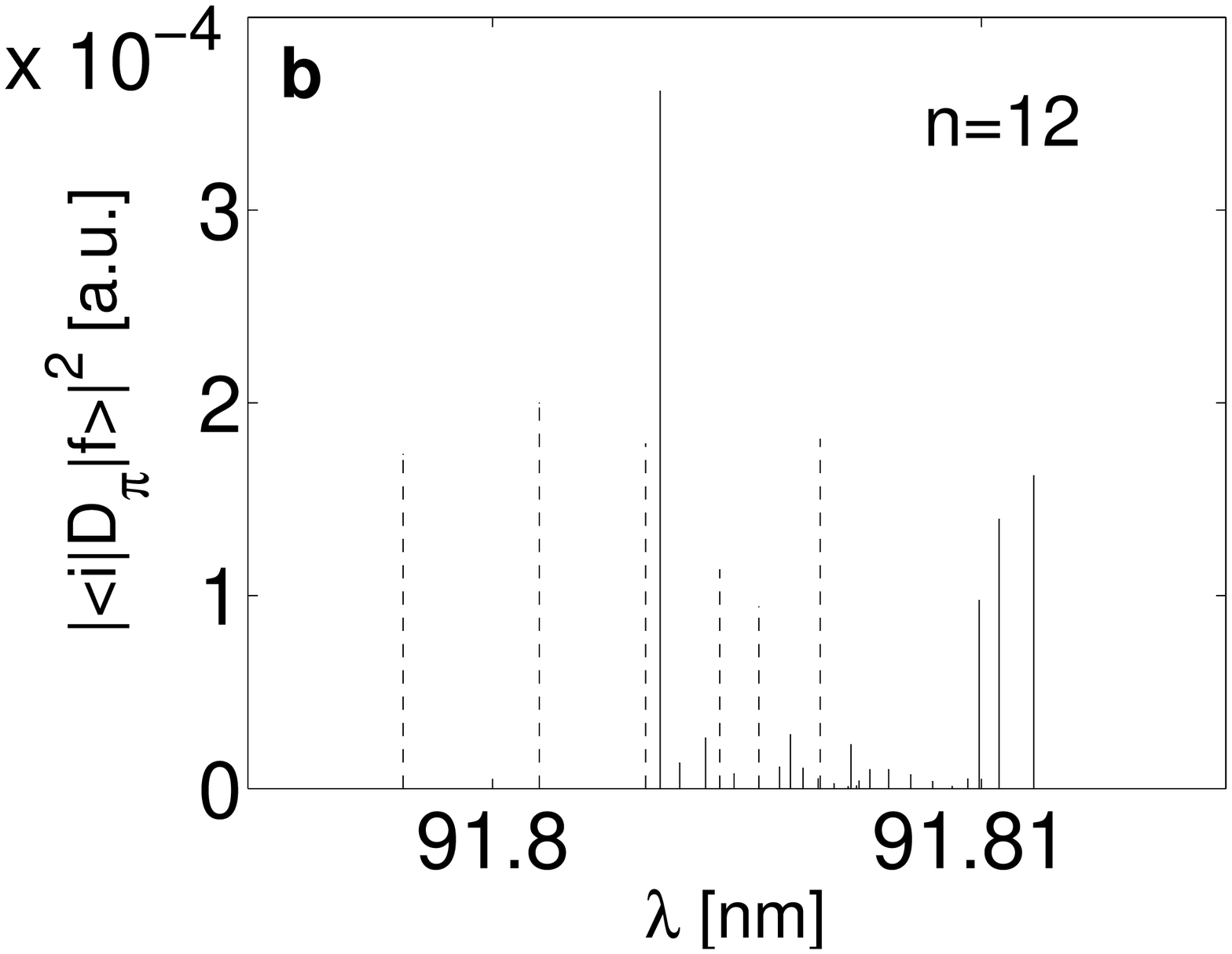}
\caption{Dipole strengths of $\triangle m=0$ $\pi-$transitions
from the ground state to excited levels. Solid lines denote
transitions in the quadrupole field whereas transitions in the
homogeneous field are indicated by dashed lines.
 \textbf{a:} The line possessing the largest wavelength belongs to
the $n=1\rightarrow 10$ transition. \textbf{b:} Magnification of
the line structure of the $n=1\rightarrow 12$
transitions.}\label{fig:lin_transition}
\end{figure}
\begin{figure}[ht]\center
\includegraphics[angle=0,width=7.3cm]{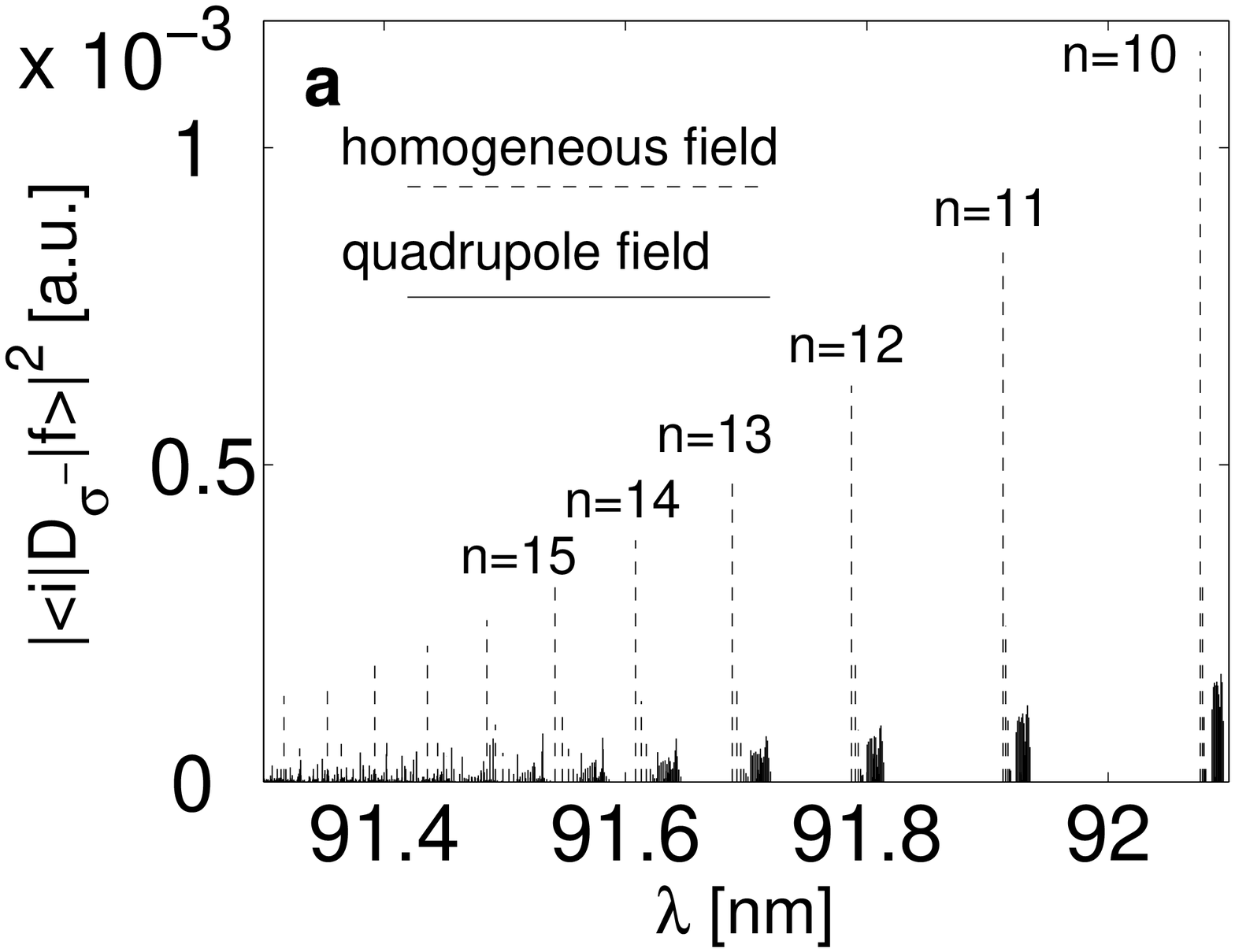}
\includegraphics[angle=0,width=7.3cm]{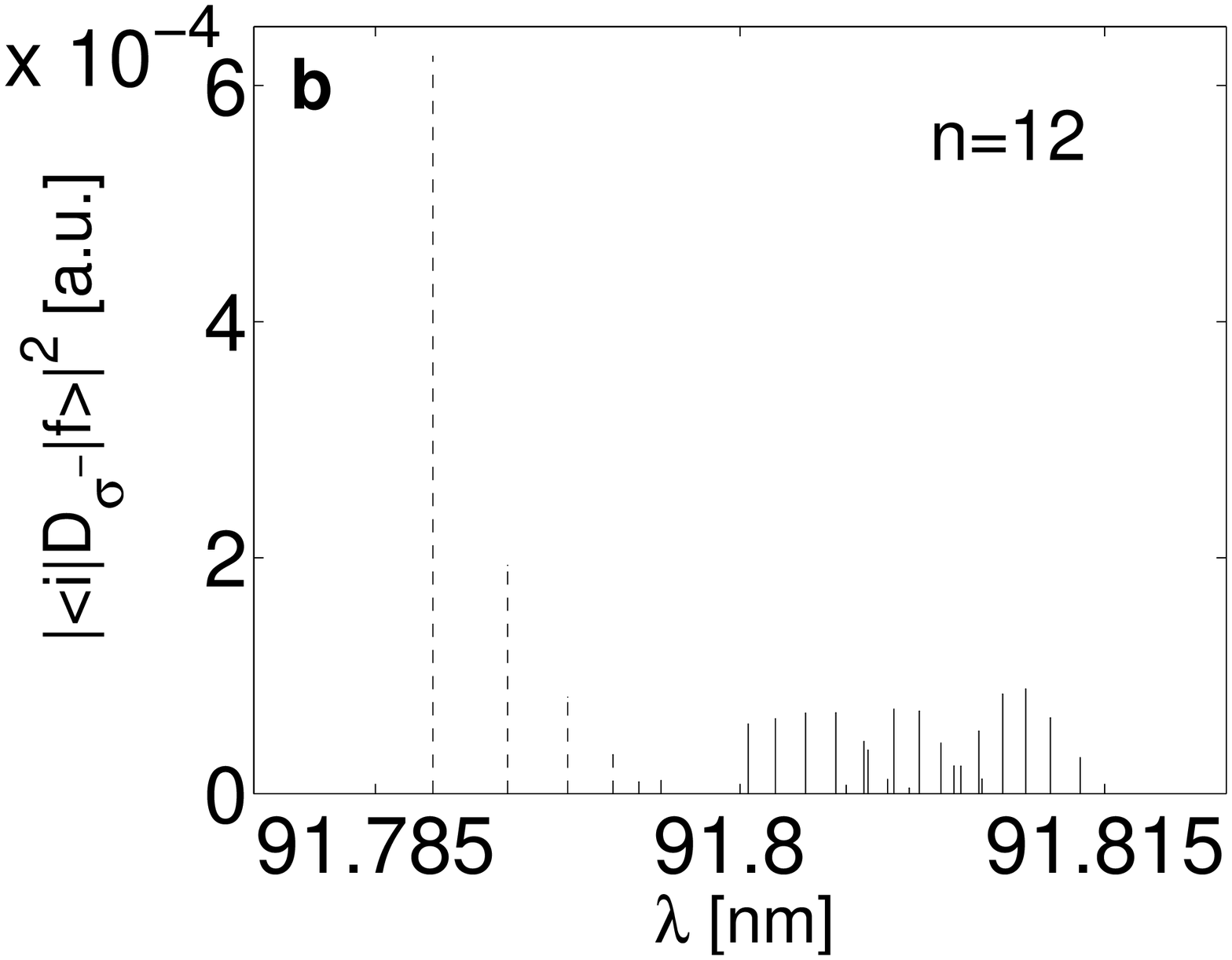}
\caption{Dipole strengths of $\triangle m=-1$ $\sigma-$
transitions ($\sigma^-$-transitions) from the ground state to
excited levels. Solid lines denote transitions in the quadrupole
field whereas transitions in the homogeneous field are indicated
by dashed lines. \textbf{a:} The line possessing the lowest
wavelength belongs to the $n=1\rightarrow 10$ transition.
\textbf{b:} Magnification of the line structure of the
$n=1\rightarrow 12$ transition.}\label{fig:circ_transition}
\end{figure}
We have calculated the dipole strengths for transitions from the
ground state ($m=\frac{1}{2}$) to excited states with
$m=\frac{1}{2}$ ($\pi$-transitions (figure
\ref{fig:lin_transition})) and $m=\frac{3}{2}$
($\sigma^-$-transitions (figure \ref{fig:circ_transition})). For
comparison we also present the transitions for an atom in a
homogeneous magnetic field indicated by dashed lines whereas
transitions in the quadrupole field are represented as solid
lines. For a detailed discussion of electromagnetic transitions in
the homogeneous field we refer the reader to
\cite{Ruder94,Clark80,Clark82}. In order to make both results
comparable we have chosen the gradient and the magnetic field
strength such that we observe approximately equal splitting of the
$n$-multiplets with increasing energy. Specifically we chose
$b=10^{-7}$ and $B=10^{-4}$. Figure \ref{fig:lin_transition}a
shows the dipole strengths for $\pi$-transitions. Here the line
possessing the largest wavelength emerges from the transition to
the $n=10$ multiplet which already exhibits significant features
of intra $n$-manifold mixing. From here up to a certain wavelength
($\approx 91.5 nm$) neighboring lines are well separated.
Transitions with smaller wavelengths involve levels showing inter
$n$-manifold mixing becoming visible by the overlapping of
neighboring groups of lines. In figure \ref{fig:lin_transition}b
we provide a higher resolution picture of the transitions to the
$n=12$-manifold in the intra $n$-mixing regime. Apparently in both
cases each main line is accompanied by a series of sub-lines. In
the quadrupole field besides the main line at $\lambda \approx
91.803 nm$ there are four major sub-lines situated at the outer
edge $\lambda \approx 91.81 nm$. We observe a number of sub-lines
almost equal in height compared to the main line for the
homogeneous field. We note that the sub-line possessing the
maximum strength always belongs to a transition in the quadrupole
field. An overall feature is the fact that a significantly lower
number of sub-lines occurs in the homogeneous field compared to
the quadrupole field:  Symmetry properties deriving from the
conserved $z$-parity give rise to additional selection rules and
lead to a reduction of the number of allowed transitions. In
figure \ref{fig:circ_transition} we present the transitions to
states $m=\frac{3}{2}$ starting from the transition to the $n=10$
multiplet. Again the change from the inter to the intra $n$-mixing
regime is observed (figure \ref{fig:circ_transition}a). Here the
threshold wavelength at which the overlapping of adjacent groups
of lines starts is also at $\lambda\approx 91.5 nm$. Compared to
the wavelengths in the homogeneous field we find the transitions
in the quadrupole field to be systematically shifted towards
larger wavelengths. In figure \ref{fig:circ_transition}b we show
the magnification of the line structure for the $n=1\rightarrow
12$ transition. In the homogeneous field each transition to a
fixed $n$-multiplet is dominated by a major line accompanied by a
group of significantly weaker sub-lines. In contrast to this we
find many transitions of equal strength in the quadrupole field.
This is in contrast to the above-discussed $\pi$-transitions. The
total number of transitions is again much larger in the quadrupole
field compared to the homogeneous field.

%
\subsection{Magnetic field induced permanent electric dipoles}\label{subsec:dipole_moments}
%
In a homogeneous magnetic field parity is a symmetry and therefore
atomic electronic eigenstates (for fixed nucleus) do not possess a
permanent electric dipole moment. Let us now investigate the
electric dipole moment of the electronic states in the quadrupole
field. Due to the selection rule given above the expectation value
of $D_{\sigma^\pm}$ in the $J_z$-eigenstates is
\begin{eqnarray}
\left<D_{\sigma^\pm}\right>=\left<E,m\right|r\sin\theta\,e^{\pm
i\phi}\left|E,m\right>=0.
\end{eqnarray}
\begin{figure}[ht]\center
\includegraphics[angle=0,width=7.3cm]{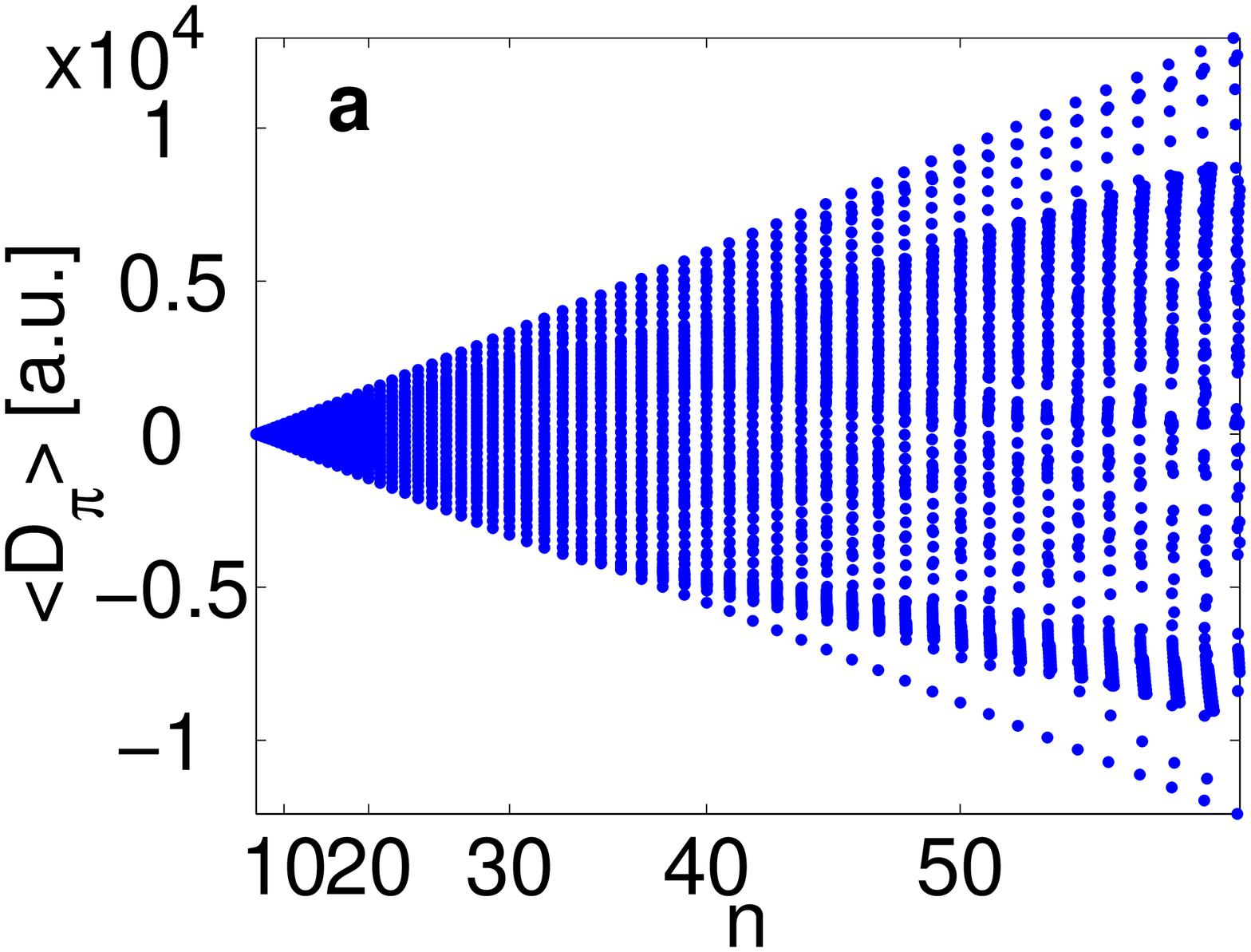}
\includegraphics[angle=0,width=7.3cm]{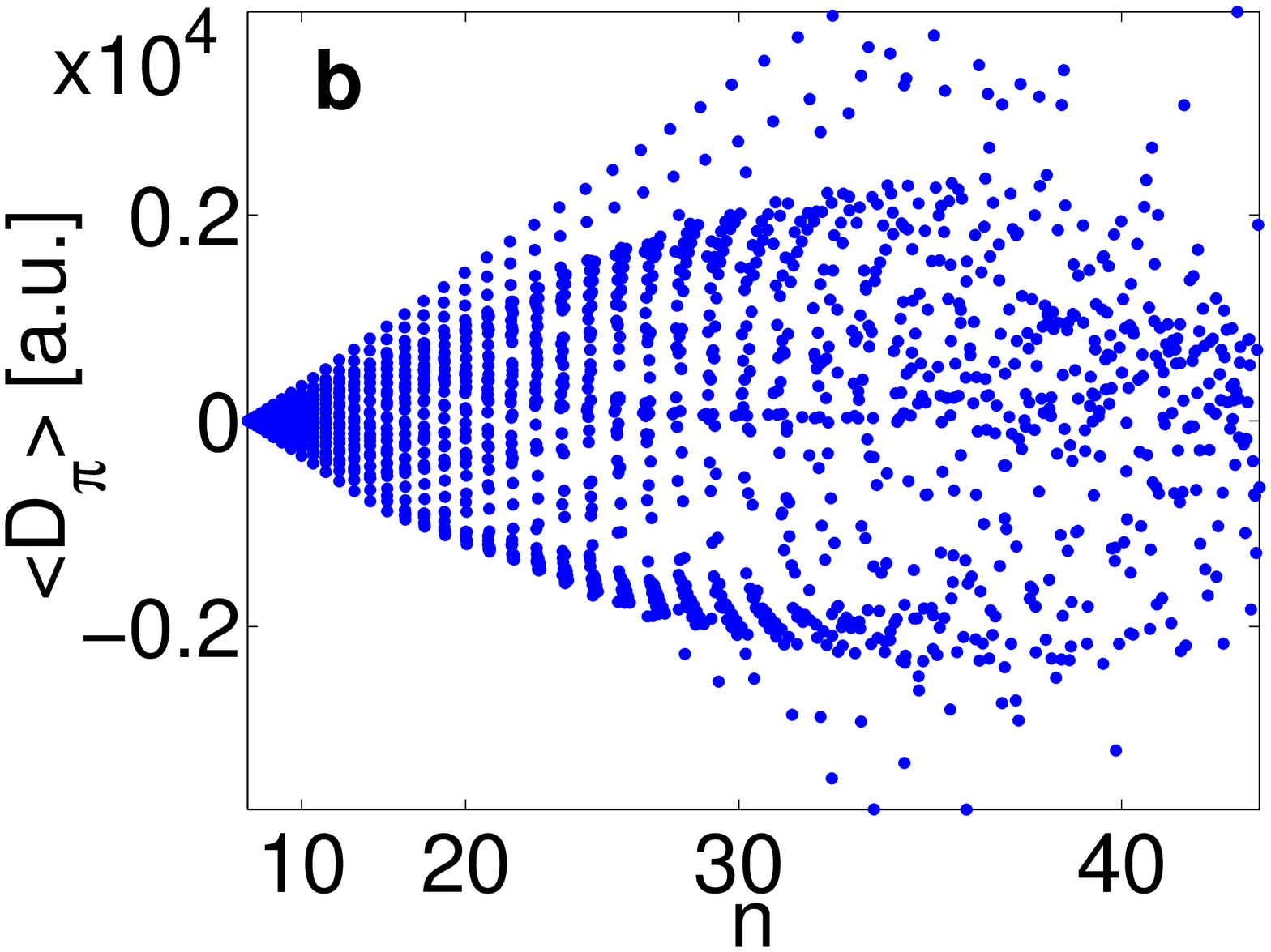}
\caption{Expectation value of the dipole operator $D_\pi$ plotted
against the principle quantum number $n$ for different gradients
(\textbf{a}: $b=10^{-10}$, \textbf{b}: $b=10^{-8}$).}\label{fig:dipolemoment}
\end{figure}
However, the expectation value of $D_\pi$ is in general non-zero.
$\left<D_\pi\right>$ is shown in figure
\ref{fig:dipolemoment}(a,b) for the two gradients $b=10^{-10}$ and
$b=10^{-8}$, respectively. For $b=10^{-10}$ the electric dipole
moments belonging to the same $n$-multiplet are arranged along
vertical lines, which is a result of the approximate degeneracy of
the energy levels. The substates for fixed $m$ of a given
$n$-multiplet exhibit different dipole momenta spreading between
two (upper and lower) bounds that depend linearly depending on
$n$. States with a large electric dipole moment emerge from
field-free states (with increasing $b$) that possess small values
for the angular momentum and vice versa. With increasing gradient
and degree of excitation the $n$-mixing starts and disturbs the
observed regular pattern. For $b=10^{-8}$ and $n>35$ the
distribution of the dipole moments becomes completely irregular.
Therefore we encounter the remarkable effect that the external
magnetic quadrupole field induces a state dependent permanent
electric dipole moment.  This is the result of the asymmetric form
of the wavefunction in the quadrupole field (see discussion in
section \ref{subsec:radial_compression} and figure
\ref{fig:low_states}). Exciting the atom from the ground to a
corresponding excited state via lasers with the corresponding
transition frequency it is therefore possible to prepare an atom
with a desired permanent electric dipole moment !

%
%
\section{Conclusion and Outlook}\label{sec:conclusion_outlook}
%
%
We have presented a detailed study of the electronic structure of
atoms in a magnetic quadrupole field. Opposite to the common
adiabatic description of the atomic center of mass motion in
inhomogeneous magnetic fields we have focussed on the internal
electronic states of the atoms exposed to the inhomogeneous field.
We have employed an effective one-particle approach including both
the coupling of the electric charge and the magnetic moment (spin)
to the field. A spinor-orbital based method to compute the
eigenfunctions of the stationary Schr\"odinger equation has been
developed and applied. We have utilized a 'Sturmian' basis set to
study several thousands of excited states for a regime of
gradients of more than $10$ orders of magnitude. Due to the
inhomogeneity of the quadrupole field the spatial and spin degrees
of freedom are coupled in a unique way. As a result the system is
invariant under a number of symmetry operations acting on both
degrees of freedom. These are the unitary symmetries related to
the conservation of the total angular momentum $J_z$ and the
discrete operation $P_\phi OP_z$ as well as the anti-unitary
generalized time reversal symmetries $TOP_z$ and $TP_\phi$. These
operations constitute a non-Abelian symmetry group which leads to
a two-fold degeneracy of each energy level. This is a remarkable
feature in particular since it occurs in the presence of the
external field. Without a field such (Kramers) degeneracies of a
spin-$\frac{1}{2}$-particle are well-known.

We have shown energy spectra up to excitation energies
corresponding to principal quantum numbers of $n\approx 60$. The
spectrum exhibits distinct characteristics for the weak,
intermediate as well as the strong gradient regime. At weak
gradients a linear splitting of the energy levels is observed. For
given $m$ the sublevels of an $n$-multiplet split symmetrically
around the zero field energy. For the homogeneous field a
splitting in only two branches is observed due to the two
energetically different orientations of the electronic spin.
Employing higher gradients adjacent $n$-manifold are still
separated but we encounter intra $n$-manifold mixing. Scaling
relations for the onset of the intra- as well as the inter
$n$-manifold mixing in a quadrupole field have been given. To
understand some general features of the effects generated by the
quadrupole field we have studied the spectrum of low excitations
in very strong (experimentally not accessible) gradients. We have
found that the ground state energy virtually stays the same up to
gradients of $b=10^{-2}$ whereas the first few excited states
already experience severe changes. The latter obviously only holds
for hydrogen. However, one way to proceed is to investigate
Rydberg states of nonhydrogenic atoms in detail. Here especially
scattering with the inner electron shell leads to modifications of
the energy spectra which can be understood by quantum defect
theory \cite{Schmelcher98,Wang91}. At least for Rydberg states
which possess a large angular momentum we do not expect
significant modifications induced by core scattering processes.

In our investigation of the electronic spin properties we found
the $S_z$-expectation values of the electronic states to form a
regular pattern at low gradients. For higher gradients and/or
higher excitations we observe a transition to an irregular, much
narrower, distribution. In order to analyze the spatial behavior
of the electronic spin we introduced the $S_z$-polarization. Due
to the unique coupling of the spatial and spin degrees of freedom
this quantity reveals a rich nodal structure. At low radii we
observe a chess-board like pattern of islands with alternating
spin polarization changing to a striped pattern at large radii.
For Rydberg states the chess board pattern changes to an irregular
network of islands exhibiting opposite spin polarizations. Within
the striped region an envelope behaviour of the spin orientation
on the angle $\theta$ is observed. To investigate this phenomenon
we have studied a Hamiltonian describing the coupling of the
electronic spin to the magnetic field, neglecting the spatial
dynamics. Analytically calculating its eigenstates we could
reproduce the above mentioned $\theta$-dependence of the
$S_z$-polarization of Rydberg-states for large radii.

We have derived selection rules for electromagnetic dipole
transitions among eigenstates to the $J_z$ as well as $TOP_z$
operator. Linearly polarized transitions take place only between
states with the same quantum number $m$ or states with different
$TOP_z$ symmetry. Furthermore we have calculated the amplitudes
for $\pi$- and $\sigma^-$-transitions emerging from the ground
states ($m=\frac{1}{2}$) to levels lying in the intra and inter
$n$-manifold mixing regime. We have compared the results with
those of a homogeneous magnetic field discovering significant
differences in the line structure. Due to additional selection
rules arising from the conservation of $z$-parity fewer
transitions are observed in a homogeneous magnetic field.
Calculating the expectation value of the dipole operator it turned
out that the electric dipole moment of the electronic states in the
quadrupole field is in general nonzero. For low degrees of excitation and low gradients
we observed an almost linear increase of the maximum dipole moment
of the $n$-multiplets. This pattern becomes increasingly distorted
when moving to higher gradients and/or a higher degree of excitation. The
non-vanishing dipole moment is a result of the spatially
asymmetric arrangement of the electronic wavefunction with respect to
the $x-y$-plane.

In order to achieve the experimental realization of the discussed
system the so called atomchip \cite{Folman02} seems to be a
promising device. Here current carrying nano-structures with
dimensions in the $\mu m$-regime can generate high gradient
quadrupole fields. At the moment gradients such as $b=10^{-10}$ or
even slightly above represent certainly the limit. However, almost
all effects discussed here are not due to the diamagnetic
interactions but have their origin in the interplay between the
Coulomb as well as the spin-spatial Zeeman terms. Therefore weak
gradients should not represent a principle obstacle to
experimentally observe the derived properties. In particular the
magnetic field-induced permanent electric dipole moments would
potentially find applications in e.g. quantum information
processing \cite{Lukin01,Jaksch99,Calarco00,Eckert02}. Exciting
the atom from the ground to a corresponding excited state via
lasers with the corresponding transition frequency it is possible
to prepare an atom with a desired permanent electric dipole
moment. In order to build a working two qubit gate one is
interested in finding ways to establish a controlled interaction
between two qubits to gain a phase shift. Considering a trapped
Rydberg state as a qubit the interaction between two of them can
be realized via dipole-dipole interaction. This interaction could
be switched on and off on demand by changing the atomic state in
the quadrupole field.

Taking into account the finite mass of the nucleus it has been
shown \cite{Avron78,Johnson83,Schmelcher94} that the center of
mass (CM) and electronic motion do not separate i.e. do not
decouple in the presence of a homogeneous magnetic field. To enter
the corresponding regime where the residual coupling becomes
important certain parameter values (excitation energy, CM energy
etc.) have to be addressed. A variety of intriguing phenomena due
to the mixing of the electronic and CM motion have been observed
consequently. Examples are the classical diffusion of the CM due
to its coupling to the chaotic internal motion
\cite{Schmelcher92}, the giant dipole states of moving atoms in
magnetic fields \cite{Dippel94} as well as the the self-ionization
process \cite{Schmelcher95a,Schmelcher95b,Melezhik00} due to
energy flow between the CM and electronic degrees of freedom. In
the present investigation we have fixed the position of the CM of
the atom at the center of the quadrupole field i.e. we have
adopted an infinitely heavy nucleus. On the other hand assuming
that an experimentally prepared atom is ultracold will certainly
minimize the CM motional effects. Nevertheless, a residual
coupling is unavoidable and its impact on the electronic structure
is, at this point, simply unknown: A full treatment of the
two-body system certainly goes beyond the scope of the present
investigation and requires both from the conceptual as well as
computational point of view major investigations. However, one
should note that the symmetries discussed here equally hold for
the moving two-body system i.e. the total angular momentum is
conserved and the unitary as well as antiunitary spin-spatial
symmetries, now applied to both particles, are also present.

\section{Acknowledgments}\label{sec:acknowledgments}

We thank Ofir Alon for fruitful discussions regarding the symmetry
properties of the atom in the magnetic quadrupole field. I.L. acknowledges
a scholarship by the Landesgraduiertenf\"orderungsgesetz of the country
Baden-W\"urttemberg.

\end{document}